\documentclass[%
 reprint,
 amsmath,amssymb,
 aps,
]{revtex4-2}
\usepackage{amsmath}
\usepackage{graphicx}
\usepackage{subcaption}
\usepackage{lastpage}
\usepackage{float}
\usepackage{fancyhdr}
\usepackage[english]{babel}
\usepackage{array}
\usepackage[T1]{fontenc}
\usepackage[usenames,dvipsnames]{xcolor}
\usepackage{hyperref}

\newcommand {\bydef}{\,\raise.07485ex\hbox{:}\kern-.025em\hbox{=}\,}
\newcommand{\Lin} {\mathbb{L}\mathtt{in}}

\newcommand {\Bc}  {\mathcal{B}}

\newcommand {\Rc}  {\mathcal{R}}
\newcommand {\Sc}  {\mathcal{S}}
\newcommand {\Tc}  {\mathcal{T}}

\newcommand {\Vc}  {\mathcal{V}}

\newcommand {\ub} {\mathbf{u}}

\newcommand {\Cb} {\mathbf{C}}

\newcommand {\Fb} {\mathbf{F}}

\newcommand {\Ib} {\mathbf{I}}

\begin{document}

\preprint{APS/123-QED}

\title{On the interplay between activity, elasticity and liquid transport\\ in self-contractile biopolymer gels}

\author{Anne Bernheim-Groswasser}
\author{Gefen Livne}
 \affiliation{Dept. Chemical Engineering, Ben-Gurion University, Israel}
 \email{bernheim@bgu.ac.il}
 \email{livneg@post.bgu.ac.il}
\author{Paola Nardinocchi}
\author{Filippo Recrosi}
\affiliation{
 Dept. Structural Engineering \& Geotechnic, Sapienza Universit\`a di Roma, Italy}
\email{paola.nardinocchi@uniroma1.it}
\email{filippo.recrosi@uniroma1.it}
\author{Luciano Teresi}
\affiliation{
 Dept. Mathematics \& Physics,  Universit\`a Roma Tre, Italy}
 \email{teresi@uniroma3.it}
%


\date{\today}

\begin{abstract}
Active gels play an important role in biology and in inspiring biomimetic active materials, due to their ability to change shape, size and  create their own morphology; the relevant mechanics behind these changes is driven by self-contraction and liquid flow.  Here, we couple contraction  and liquid flow  within a nonlinear mechanical model of an active gel disc to discuss how contraction dynamics inherits length scales which are typical of the liquid flow processes. 
The cylindrically symmetric model we present, which recapitulate our previous theoretical modeling in its basic lines, reveals that when also liquid flow is taken into account, the aspect ratio of the disc is not the only geometrical parameter which characterizes the contraction dynamics of the gel. The analyses we present  provide important insights into the dependence of contraction dynamics on geometry and allow  to make some progress in designing materials which can be adapted for different applications in soft robotics.
\end{abstract}

\maketitle


\section{\label{sec:Intro}Introduction}

Self-contractile active gels are usually generated by polymerizing actin in the presence of cross-linkers  and clusters of myosin as molecular motors\cite{Bendix:2008,Koenderink:2009,Schuppler:2016,Bernheim:2018,Ideses:2018}. 
Mechanics of active gels present interesting characteristics: self-contractions  generate internal stresses and stiffen the material, so driving the network into a highly nonlinear, stiffened regime \cite{Koenderink:2009}; morphing from flat to curved geometries can be expected when thin discs of active gels are considered \cite{Ideses:2018}; boundaries affect morphing \cite{Schuppler:2016}. 
\\
The first  models \cite{MacKintosh:2008,Banerjee:2011,Ronceray:2016} of these materials were based on a physical description of the contraction dynamics within the framework of active generalized hydrodynamics: transient force dipoles are generated by myosin pulling on actin chains and  creating active contractile stresses. These models are very accurate in modeling the contraction dynamics, looking at the network mesh scale, and less interested in coupling that dynamics with the nonlinear mechanics of active gels, which is also strongly affected by liquid flow \cite{Ideses:2018}.\\
%
%
Recently, the mechanics of active gels have been at the centre of a few  theoretical studies, set within the framework of nonlinear mechanics, where the interactions between elastic stresses and liquid flow have been investigated \cite{Curatolo:2017_M, Bacca:2019,Curatolo:2019,Curatolo:2021,Elsevier:2022}. 
In \cite{Bacca:2019}, a dynamic cross-linking mechanism is introduced to take into account the active behaviour of the gel. It drives an evolution of the mechanical stiffness of the polymeric network and an increase of the strain energy. The approach exploited  in \cite{Curatolo:2017_M,Curatolo:2019,Curatolo:2021,Elsevier:2022} by some of the authors is quite different: the activity in the gel is modeled as an external remodeling force that drives the microscopic reorganization of the network due to activity and competes with the passive deformation of the gel due to elasticity and liquid flow \footnote{See Ref.\citenum{Turzi:2017}, where a similar point of view has been used to model  active nematic gels.}. Network remodeling drives both the evolution of the mechanical stiffness of the polymer and the chain shortening, which are two of the main mechanisms  \cite{Ideses:2018,Banerjee:2011} evidenced in the experiments \cite{Ideses:2018}.\\ 
In the present work, we start from that approach \cite{Elsevier:2022} to focus on the interactions between activity, elasticity and liquid diffusion in active gels, which are largely unexplored. 
The variety of of phenomena to be understood is wide, and robust macroscopic models  of contractile networks can inspire further experiments to improve the control  of  the active characteristics of the gel and of  its relevant mechanics.\\ 
The point we discuss here is about the  competitive roles of contraction  and liquid flow in driving the mechanics of the active gel. We refer to a specific problem, whose analysis has been inspired by the work presented in Ref.\citenum{Ideses:2018}, where the contraction dynamics of an active gel disc, whose geometry is defined by  radius and thickness, has been followed and described with great details. Through the analysis of the problem, we'll show  how: (i) gel dynamics inherits length scales which are typical of the liquid flow processes; (ii) two different regimes characterize the dynamics of the disc, which can be ascribed to gel contraction and to liquid flow; (iii) the gel dimensions' aspect ratio (radius to thickness) impact on the gel dynamics and affects also stress distribution.\\
The model is presented in Sec. II and III. Sec. IV  describes the equilibrium states of the active gel and  Sec. IV deals with contraction dynamics.
\section{Active volume and polymer fraction}
\label{DIC}
Differently from passive polymer gels, active gels have the ability to remodel their mesh by self-contractions. The key elements of our model of active gel are here contrasted with the standard Flory-Rehner model of passive gels, which is at the bases of the stress-diffusion theories describing the chemo-mechanical interactions in swollen gels \cite{Doi:2009,Chester:2010,JMPS:2013,Fujine:2015,Curatolo:2018_SM,Curatolo:2021}.\\
A key variable in the Flory-Rehner model is the polymer fraction $\phi$, defined as the ratio between the
volume of the polymer $V_p$ and the total volume $V$: 
\begin{equation}\label{volfrac}
\phi=\frac{V_p}{V},
\quad\textrm{with}\quad
V=V_p+V_s,
\end{equation}
where $V_s$ is the volume of solvent content.
This formula is based on the assumption that a given mass of polymer occupies a constant volume $V_p$, be it dry or not; thus, any volume increase must be entirely due to the solvent volume $V_s$. 
Moreover, the Flory-Rehner model assumes that the polymer chains are not stretched at dry state, and that it is the solvent absorption that stretches these chains. Equilibrium is
given by a balance between the elastic energy, which prefers unstretched chains, and the mixing energy that favours swelling and thus requires stretching to accommodate more solvent.

The active gel model removes the assumption of constant polymer volume, and considers the volume that can be occupied by a given mass of dry polymer as an additional state variable, named \textit{active volume} $V_a$.
The volume $V_a$ can vary because of a change of the mean free-length of the polymer chains,
that is, of the average mesh size $\xi_a$ measured at dry conditions; thus, $V_a$ can be considered as
a coarse-grained modeling of the microscopic arrangement of the polymer chains. It turns out that
a change of $V_a$ also describes a change  of the effective stiffness of the gel. 
For the active gel model, the polymer fraction is measured by
\begin{equation}\label{volfrac2}
\phi=\frac{V_a}{V},
\quad\textrm{with}\quad
V=V_a+V_s.
\end{equation}
The hypothesis that the polymer chains are not stretched at dry state is maintained; thus, the thermodynamical equilibrium is still a consequence of the balance between elastic energy and mixing energy.
The new formula $\phi=V_a/(V_a+V_s)$ describes interactions between activity and solvent content. For example, we might have the same polymer fraction $\phi$ with different pairs $V_a$, $V_s$:
\begin{equation}\label{polyfrac}
   \phi=\frac{V_{ao}}{V_{ao}+V_{so}}=\frac{V_{a1}}{V_{a1}+V_{s1}}
\quad\Rightarrow\quad
\frac{V_{a1}}{V_{ao}}=\frac{V_{s1}}{V_{so}}\,,
\end{equation}
as $1/\phi = 1 + V_{so}/V_{ao}=1 + V_{s1}/V_{a1}$.
From (\ref{polyfrac}), it follows that a contraction of
the polymer network yields a proportional reduction of its solvent content. For very soft gels, as is our case, $\phi$ can be very small and a small volume contraction of $V_a$ can yield a huge expulsion of solvent volume $V_s$.
As example, by assuming $V_{ao}=1$ mm$^3$ and $V_{so}=1000$ mm$^3$, we have $\phi=1/1001$;
a contraction that halves the polymer volume yields $V_{a1}=0.5$ mm$^3$ and $V_{s1}=500$ mm$^3$.
Following our example, the average mesh size $\xi_a$ corresponding to the two active volumes 
$V_{ao}$ and $V_{a1}=V_{ao}/2$ scales as $\xi_{a1}/\xi_{ao}=(1/2)^{1/3}\simeq 0.8$. It is worth remembering that $\xi_a$ is the mesh size of the unstretched chains, which determines the so-called  spontaneous metric of the network, whereas the actual mesh size
$\xi$ is related to the actual swollen volume and determines the current metric of the network: $\xi\propto (V_a+V_s)^{1/3}$. 
Both $\xi_a$ and $\xi$ may be very different from the reference mesh size $\xi_d$ of the dry polymer, due to activity and liquid flow, see figure \ref{fig:2}.
\\
In the mathematical model, the microscopic arrangement of the polymer chains (from now on, the \textit{remodeling}) is driven by a new evolution equation with its own source term, which is represented by the external remodeling force that maintains the system steady, or that drives it out of equilibrium \cite{Prost:2015}. It affects the solvent flow in the gel: a contraction of the polymer mesh yields a liquid flow towards the boundary of the body, favouring its release. Indeed, as we shall shortly review in the following, the polymer fraction $\phi$ depends on $V_a$ through a balance equation of solvent concentration, driven by a Flory-Rehner thermodynamics, which is so affected by gel activity.

%
%
\section{Stresses, liquid fluxes and self-contractions}
\label{sef}
\begin{figure}[h]
  \includegraphics[height=4.5cm]{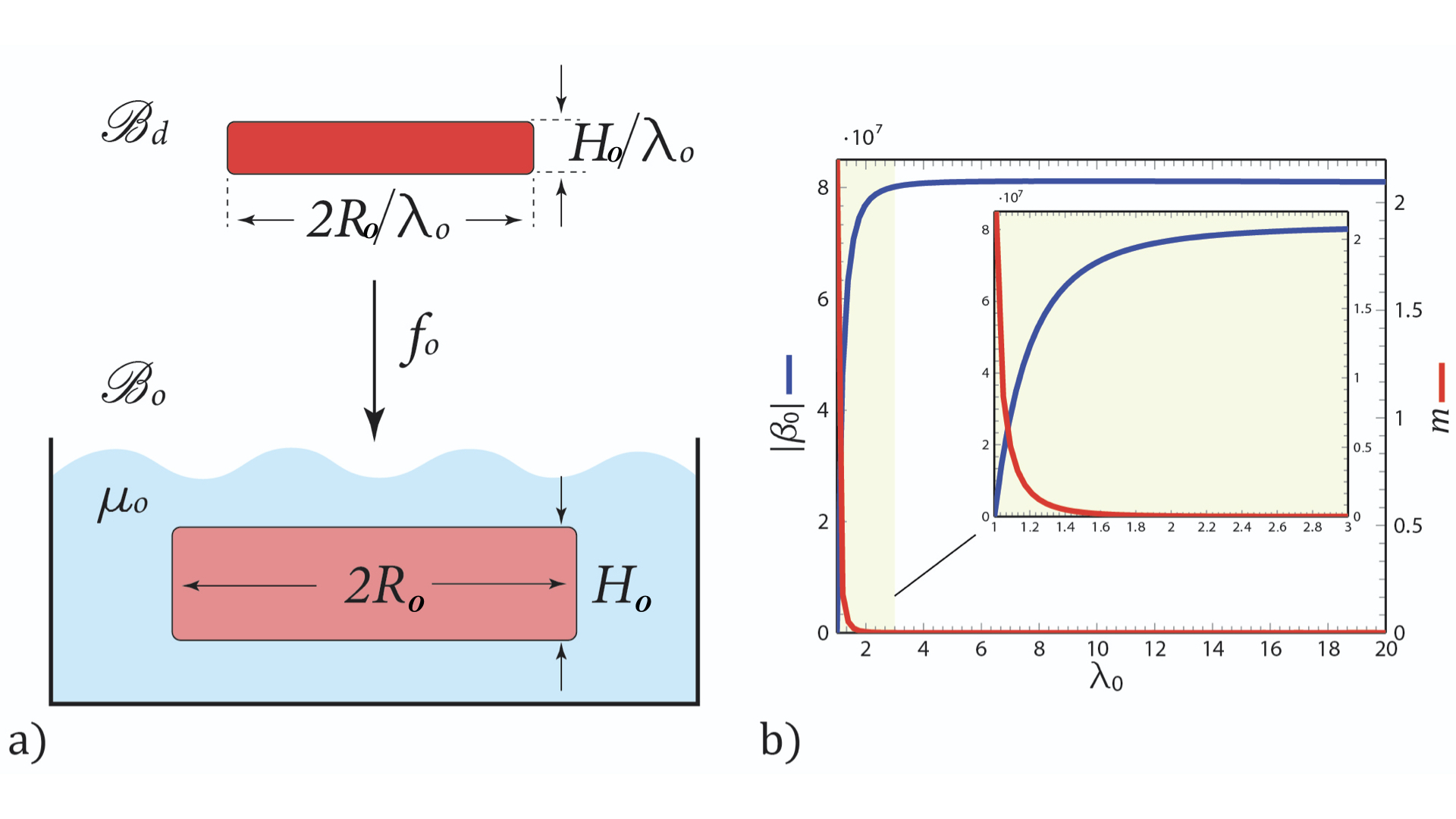}
  \caption{(a) Dry configuration $\Bc_d$ and swollen initial one $\Bc_o$: $H_o$ and $2R_o$ are thickness and diameter of $\Bc_o$ when the bath has chemical potential $\mu_o$. (b) Free-swelling stretch $\lambda_o$ versus the energy ratio $m=G\Omega/RT$ at $\mu_o=0$ (red) and versus the bulk source $\beta_o$ (blue) needed to maintain $\lambda_o$
in  thermodynamical equilibrium; the lower the ratio $m$ (that is, mixing energy larger than the elastic one), the higher the bulk source $\beta_o$.}
  \label{fig:1}
\end{figure}

%
The active gel model is formulated  in the framework of 3D continuum physics, see \cite{Curatolo:2017_M,Curatolo:2019} for details, which allow to set up initial-boundary value problems well suited to describe real experiments. Inspired by the experiments in \cite{Ideses:2018}, we consider a disc-like continuum body: at the initial time, it is a fully swollen, flat gel disc $\Bc_o$, having radius $R_o$ and thickness $H_o$, which is  $\lambda_o$ times larger then the corresponding dry disc
$\Bc_d$ (see figure \ref{fig:1}, panel a).\\ 
%
%

%

%
The region $\Bc_d$ is assumed as the reference configuration of the active gel disc and the mathematical model describes the state of the gel by using three state variables:
the solvent concentration per unit of dry volume $c_d:\Bc_d\times\Tc\to\Rc$
($[c_d]=$mol/m$^3$);
the mechanical displacement $\ub:\Bc_d\times\Tc\to\Vc$ ($[\ub]=$m);
the active contractions $\Fb_a:\Bc_d\times\Tc\to\Lin$
($[\Fb_a]=$1), usually called \textit{remodeling tensor}.
Here, $\Rc$, $\Vc$, and $\Lin$ denote a scalar, a vector and a tensor, respectively; $\Tc$ is the time interval under study (see \cite{Curatolo:2019,Curatolo:2021} for details).

\begin{figure}[h]
\centering
  \includegraphics[height=4.5cm]{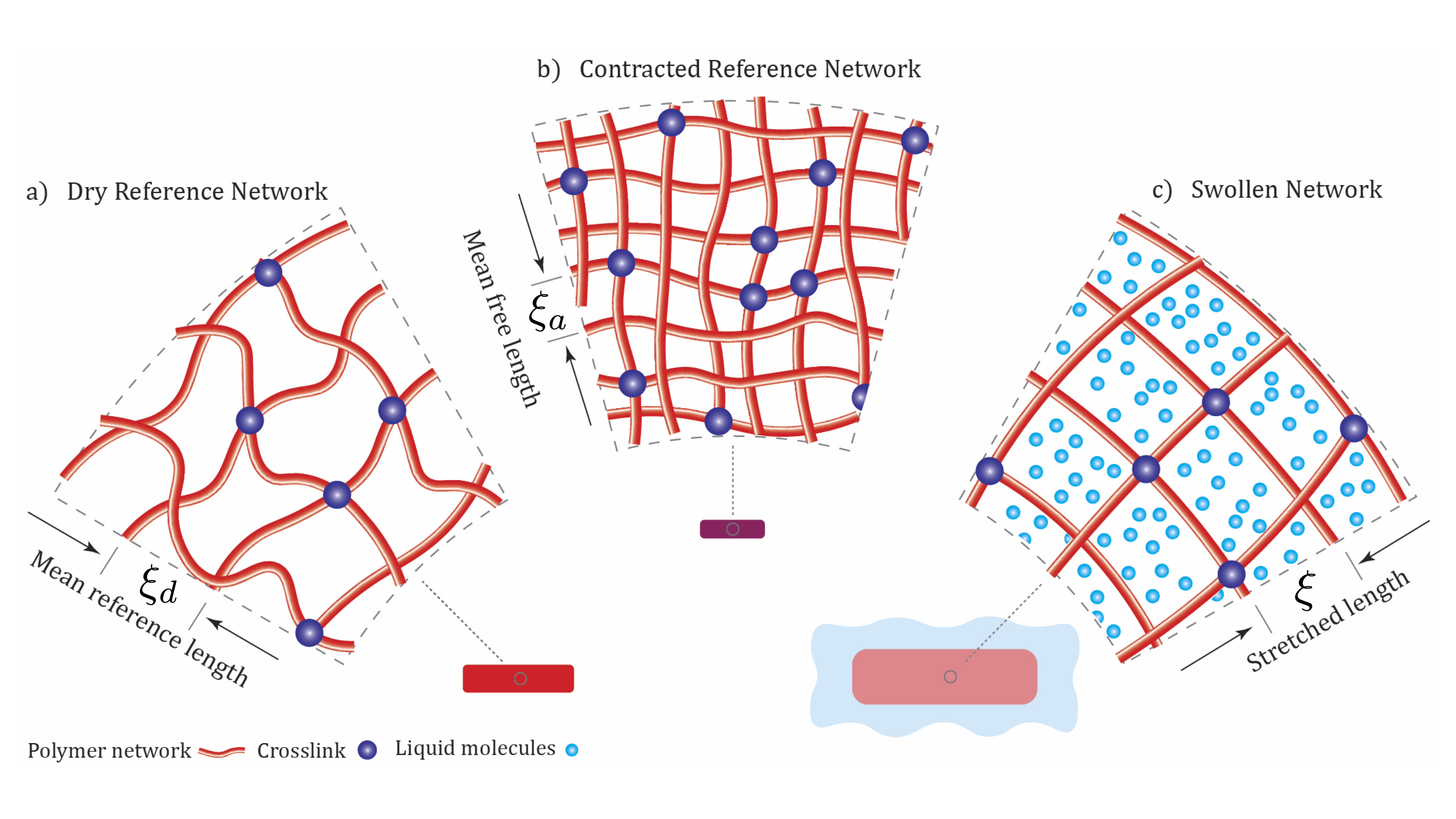}
  \caption{The characteristic states of an active gel:
  the three cartoons might be considered as representative volume elements. 
  (a) Dry-reference meshwork (red) of size $\xi_d$ with crosslinks (blue dots). 
  (b) Dry-contracted meshwork: mesh size $\xi_a$ is reduced with respect to $\xi_d$, and crosslink density is 
  higher; the polymer chains are considered unstretched. (c) Swollen meshwork: liquid molecules (light blue dots) swell the dry-contracted meshwork: the elastic energy is proportional to the stretch $\xi/\xi_a$ between the contracted mesh and the swollen one.}
  \label{fig:2}
\end{figure}

Solvent concentration $c_d$ and displacement $\ub$
are the standard state variables based on Flory-Rehner model;
the active contraction $\Fb_a$ is the new variable used
to describe active gels. The tensor $\Fb_a$ is the 3D equivalent of the volume $V_a$ mentioned in the previous section: it describes not only the change in volume, but also the macroscopic changes in length and angles of the polymeric network due to self-contractions (see figure \ref{fig:2}). The time-dependent symmetric tensor field $\Cb_a=\Fb_a^T\Fb_a$ accounts for the reduction of the free length of the
polymer chains, due to self-contraction, and describes the spontaneous metric of the gel. 

Given the deformation gradient $\Fb=\Ib+\nabla\ub$, 
the key relations (\ref{volfrac2}) are now represented in 
terms of Jacobian determinants
    \begin{equation}\label{volC}
    \phi=\frac{J_a}{J},
    \quad\textrm{with}\quad
    J=\textrm{det}\,\Fb=J_a + \Omega\,c_d,
    \quad
    J_a=\textrm{det}\,\Fb_a\,;
    \end{equation}
it holds $\xi_a/\xi_d\simeq J_a^{1/3}$ and $\xi/\xi_d\simeq J^{1/3}$. Equations (\ref{volC}) imply that any actual volume
 change $J$ is the sum of a volume change of the active mesh $J_a$, plus the volume of the solvent
 $\Omega\,c_d$. Polymer fraction $\phi$ delivers a measure of the gel density, which increases when solvent content decreases and depends on the volume change of the active mesh, as equations \eqref{volC} indicate.
\\
The deformation of the actual mesh with respect to the
unstretched one is measured by $\Fb_e=\Fb\,\Fb_a^{-1}$,
called elastic deformation and the symmetric tensor field $\Cb_e=\Fb_e^T\Fb_e$ describes the so-called  elastic metric, which affects stresses distribution in the network.

%
\subsection{Model equations under cylindrical symmetry}\label{cylsym}
We  exploit the cylindrical symmetry that greatly simplifies the evolution equations of the problem;
thus, the reference disc $\Bc_d$ is represented by its vertical cross section $\Sc_d$ spanned by the radial coordinate $r\in(0,R_d)$ and the vertical one $z\in(0,H_d)$.
With this, the displacement $\ub$ has two non trivial components: the radial  $u$ and the vertical $w$ component;
within the class of remodeling tensors $\Fb_a$
which are
cylindrically symmetric, we  choose a diagonal one $\Fb_a=\textrm{diag}(\gamma_r,\gamma_\theta,\gamma_z)$.


Hence, the state variables  of the problem are reduced to the following six scalar fields:
the solvent concentration $c_d$,
the two displacements $(u, w)$, and the three contractions $(\gamma_r,\gamma_\theta,\gamma_z)$; each field is a function of the coordinates $(r,z)$ and the time $\tau$.
Moreover, we assume that the derivatives $u,_z$ and 
$w,_r$ can be neglected; it follows that 
the deformation gradient $\Fb$ simplifies to 
$\Fb=\textrm{diag}(\lambda_r,\lambda_\theta,\lambda_z)$ with the radial, hoop and vertical deformations defined as
\begin{equation}
\lambda_r=1+u,_r,
\quad
\lambda_\theta=1+u/r,
\quad
\lambda_z=1+w,_z\,,    
\end{equation}
respectively. Under the symmetry assumption, the volumetric constraint \eqref{volC} takes the form
\begin{equation}\label{volCs}
\lambda_r\lambda_\theta\lambda_z=1+\Omega\,c_d\,.
\end{equation}
The active chemo-mechanical state of the active gel is ruled by a set of three balance equations, which can be rationally derived from basic principles \cite{Curatolo:2017_M}:
balance of solvent content, of forces, and of
remodeling forces.
The  first two balance equations, under the cylindrical symmetry hypotheses, reduce to the following three scalar equations
\begin{equation}\label{bala1}
\begin{array}{l}
-\dot c_d = h_{r,r} + \dfrac{h_r}{r} + h_{z,z}\,,\\[3mm]
s_{r,r} + \dfrac{s_{r}-s_{\theta}}{r} =0\,, \\[3mm]
s_{z,z}  =0\,.
\end{array}
\end{equation}
In equations \eqref{bala1},  $h_r$ and $h_z$ are the radial and vertical components of the solvent flux, whereas
$s_r$, $s_\theta$  and  $s_z$ are the radial, hoop and vertical components of the reference stress (also called Piola-Kirchhoff stress).\\
Flux, chemical potential $\mu$  and stresses
are related to the stretches $\lambda_i$ and the contractions $\gamma_i$ ($i=r,\theta,z$),
by constitutive equations, whose derivation is fully described in many texts and papers (see \cite{Gurtin:2010,Hong:2008,JMPS:2013}). Shortly, liquid transport in the elastic solid is described by a kinetic law, based on the assumption that the liquid molecules diffuse in the gel and the coefficients of diffusion can be different in the radial and vertical direction but independent of the deformation and the concentration. In the end, the liquid flux is related to the gradient of the chemical potential by the following equations
\begin{equation}\label{fluxes}
    h_r=-\dfrac{D_r \, c_d}{R\,T\,\lambda_r^2}\,\mu,_r\quad\textrm{and}\quad
    h_z=-\dfrac{D_z \, c_d}{R\,T\,\lambda_z^2}\,\mu,_z\,,
\end{equation}
where  $D_r$ and  $D_z$  are the coefficients of diffusion
in the radial and vertical direction, $R$ and $T$  are the gas constant and the temperature, respectively, and $\mu$ is the chemical potential of the solvent in the gel: 
\begin{equation}\label{chemp}
    \mu=R\,T\,g(J_e) +\Omega\, p\,,
\end{equation}
with 
\begin{equation}\label{gJe}
g(J_e)=\left[ \log\left(\frac{J_e-1}{J_e}\right) + \frac{\chi+J_e}{J_e^2} \right]\,,\quad J_e=\textrm{det}\,\Fb_e=\frac{J}{J_a}\,.
\end{equation}
Therein, $\Omega$ is the molar volume of the liquid ($[\Omega]=$ m$^3$/mol)  and  $\chi$ is the non dimensional dis-affinity parameter\cite{Doi:2009}.
The pressure field $p$ is is the Lagrangian multiplier of the constraint $ J=J_a+\Omega\,c_d$ (equation \eqref{volC}).
Finally, the stresses are given by
\begin{eqnarray}
    s_r&=&G\,\lambda_r\,\frac{\gamma_\theta\gamma_z}{\gamma_r} - p\,\lambda_\theta\lambda_z\,,\nonumber\\
    s_\theta&=&G\,\lambda_\theta\frac{\gamma_r\gamma_z}{\gamma_\theta}-p \,\lambda_r\,\lambda_z\,,\label{sigma}\\
    s_z&=&G\,\lambda_z\frac{\gamma_r\gamma_\theta}{\gamma_z}-p \,\lambda_r\,\lambda_\theta\,,\nonumber
\end{eqnarray}
where $G$ is the shear modulus of the dry polymer network ($[G]=$J/m$^3$). The (actual) Cauchy stresses are: $\sigma_r=s_r/\lambda_\theta\lambda_z$, $\sigma_\theta=s_\theta/\lambda_r\lambda_z$ and $\sigma_z=s_z/\lambda_\theta\lambda_r$.\\
%
%
The third balance equation, which describes the time evolution of the spontaneous metric delivered by the self-contractions $\gamma_i$, 
reduces to three scalar equations \footnote{See \cite{Elsevier:2022} for a detailed derivation of the equations below.}:
\begin{equation}\label{rem}
\begin{array}{l}
\dfrac{\dot\gamma_r}{\gamma_r}=
  \dfrac{1}{\eta_r}\,(\beta_r-E_r)\,,\\[4mm]
\dfrac{\dot\gamma_\theta}{\gamma_\theta}=
\dfrac{1}{\eta_\theta}\,(\beta_\theta-E_\theta)\,,\\[4mm]
\dfrac{\dot\gamma_z}{\gamma_z}=
\dfrac{1}{\eta_z}\,(\beta_z-E_z)\,.
\end{array}
\end{equation}
The evolution of the self-contractions $\gamma_i$ is driven  by the differences $(\beta_i-E_i)$ $(i=r,\theta,z)$. Therein, $\beta_i$ describes the effect of molecular motors on the mesh, is a control parameter of the model and will be denoted as \textit{active stress} from now on. On the other hand,  the three functions  $E_i$ are the components of the Eshelby tensor, which is completely determined by the elasto-chemical state of the gel through the Flory-Rehner free-energy and the stress state in the gel as:
\begin{equation}\label{esh-r}
E_i =  e_y - J\,\sigma_i\,,\quad (i=r,\theta,z)
\end{equation}
with 
\begin{equation}\label{esh_stress-free}
    e_y = \frac{R\, T}{\Omega}\,J_a\, \bigl(f_c(J_e)  + m\,f_e(\Cb_e)\bigr) -c_d\,\mu\,.
\end{equation}
Therein, $f_c$ and $f_e$ are the dimensionless mixing and elastic free-energy:
\begin{eqnarray}
    f_c(J_e) &=& (J_e-1)\textrm{log}(1-\frac{1}{J_e}) + \chi (1-\frac{1}{J_e})\,,\nonumber\\
    f_e(\Cb_e) &=&\frac{1}{2}(\textrm{tr}\Cb_e-3)\,.\label{frees}
\end{eqnarray}
So, the equations \eqref{rem}-\eqref{frees} show as the interplay between activity, elasticity and liquid transport depends on the effective stresses $(\beta_i-E_i)$ and on the frictions  $(\eta_r,\eta_\theta, \eta_z)$ of the mesh, that is, the resistance of the mesh to remodel, in the three-directions. Frictions bring in the model one or more characteristic times, which affect the mesh remodeling and though it the whole process. Large frictions yield small contraction time rates, under the same effective stresses.\\
As a first work hypothesis, we assume $\beta_i$ to be uniform and isotropic: $\beta_r=\beta_\theta=\beta_z=\beta$. 
We also assume that the disc is not constrained, nor loaded and,
as chemical boundary conditions, we assume that all the disc boundary is permeable and  chemical equilibrium holds at the boundary, that is,
\begin{equation}\label{bcc}
\mu=\mu_e\quad\textrm{on}\quad\partial\Sc_d\,,
\end{equation}
where $\mu_e$ is the difference between the chemical potential of the  bath and that of pure water ($\mu_e=0$ corresponds to a pure water bath).
Finally, the  initial conditions for the displacements $u, w$, the concentration $c_d$ and the contractions $\gamma_i$ 
($i=r,\theta,z$) are the following:
\begin{equation}\label{ic}
u = (\lambda_o-1)\,r,
\quad
w = (\lambda_o-1)\,z,
\quad c_d=c_{do},
\quad \gamma_i=1\,.
\end{equation}
It means that the deformation $f_o$ from the reference region $\Bc_d$ to the initial region $\Bc_o$ is $f_o(X)=\lambda_o\,X$ for any $X\in\Bc_d$ (see figure \ref{fig:1}).\\
\subsection{Details of Finite Element Analysis}\label{FEA}
Equations \eqref{volCs}, \eqref{bala1} and \eqref{rem}, together with the boundary \eqref{bcc} and initial \eqref{ic} conditions, are rewritten in a weak form and implemented in the software COMSOL Multiphyisics
by using the Weak-Form physics interface. The calculus domain is the rectangular domain $\Sc_d$, which  is meshed with triangular elements whose maximum mesh size is  $H_d/10$, yielding about 200K dofs. Lagrangian polynomials are used as shape functions: polynomials of order 4 for the displacement and the solvent concentration, of order 3
for the volumetric constraint, of order 2 for the
boundary conditions (also implemented in weak form)
and of order 1 for the remodeling variables. The whole set of coupled equations are solved by using  the Newton method with variable damping, as nonlinear solver, the direct solver Pardiso as  linear solver is the direct solver Pardiso and the BDF method with order 1-2 as time dependent solver.\\
 As non linear method, it is used the Newton method with variable damping; the linear solver is the direct solver Pardiso, while the time dependent solver uses the BDF method with order 1-2.
The time-dependent analysis starts at the initial state $\Bc_o$ and stops at a final equilibrium state $\Bc_1$, which is pre-selected, as we'll discuss in the next section. 
\section{Initial and final equilibrium states}
\label{IFS}
The definition of the steady states where contraction dynamics and liquid transport start and finish is an important issue. Here, we get some data  on the conditions of the gel discs at the initial and final states  in the experiments which have inspired us \cite{Ideses:2018}, and  reproduce those conditions in the numerical model.\\
Firstly, we define a steady state as a solution of the balance equations
(\ref{bala1}), (\ref{rem}) with $\dot c_d=0$ 
and $\dot\gamma_i=0$ ($i=r,\theta,z$). Such a state is
controlled by the pair $(\mu_e,\beta$), that is by the  conditions
\begin{equation}\label{ssfs}
    \mu=\mu_e\quad\textrm{and}\quad E_i=\beta \quad (i=r,\theta,z).
\end{equation}
We study the contraction dynamics between the initial steady state $\Bc_o$, which is represented by a black dot in the diagram of figure \ref{fig:4}, and a final
steady state $\Bc_1$ (red or blue dots in the diagram of figure \ref{fig:4}), corresponding at a time $\tau=\tau_1$.\\
We assume that at the steady states $\Fb$ and $\Fb_a$ are
uniform and spherical, that is $\Fb=\lambda\,\Ib$, $\Fb_a=\gamma\,\Ib$, and that initial and final states are stress-free. With this, equations \eqref{sigma}, \eqref{chemp} and \eqref{esh-r} deliver a representation form for both the chemical potential and the Eshelby components, in terms of $J_a$ and $J$: $\mu=\mu(J/J_a)=\mu(J_e)$ and $E_i=e_y(J_a,J_e)$.
Equations \eqref{ssfs} deliver the relationships between the values of $J_a$ and $J$ at the initial and final states and   the pair $(\mu_e,\beta$) which determines those values:
\begin{equation}\label{ssfs2}
    \mu_e=\mu(J_e)
    \quad\textrm{and}\quad 
    \beta=e_y(J_a,J_e)\,.
\end{equation}
In the following, we adopt the following notation: $J_o$ and $J_1$ denote the values of $J$ at $\Bc_o$ and $\Bc_1$, and the same we do for all the other quantities.\\
The evolution of the system from $\Bc_o$ to $\Bc_1$, that is, the contraction-liquid transport dynamics, is triggered by defining
time laws for the two controls,  which have both a characteristic evolution dynamics. For the motors, the characteristic time is dependent on the binding/unbinding kinetics of the motors to the actin filaments. For the chemical potential, the characteristic time reflects the mixing kinetic of possibly free biopolymer chains and the liquid in the bath. We set
\begin{equation}\label{step}
\begin{array}{l}
\mu_e=\mu_e(\tau)=\mu_o
                 +(\mu_1-\mu_o)\,\textrm{s}(\tau/\tau_{\mu})\,, \\[2mm]
\beta=\beta(\tau)=\beta_o
                 +(\beta_1-\beta_o)\,\textrm{s}(\tau/\tau_{\beta})\,, 
\end{array}
\end{equation}
where $\textrm{s}(\cdot)$ is a smoothed step function running from
$0$ to $1$ in the interval $(0,1)$ and $\tau_\mu$ and $\tau_\beta$ (both less than $\tau_1$) are the characteristic time of the controls (see Table \ref{tab:1}). 
Thus, $\beta(0)=\beta_o$,
and $\beta(\tau_\beta)=\beta_1$, and analogously for $\mu_e$.
\subsubsection{Material parameters}
\begin{table}[h]
\caption{\label{tab:1} Material and geometrical parameters \label{tab:para}}
\centering
\begin{tabular}{ll}
\hline
shear modulus	& $G  = 135$ Pa\\[1mm]
\hline
Flory parameter	& $\chi  = 0.4$ \\[1mm]
\hline
water molar volume	& $\Omega =1.8e-5$ m$^3$/mol\\[1mm]
\hline
temperature     & $T  = 293$ K \\[1mm]
\hline
energy ratio  & $m  = G\,\Omega / R\,T = 1e-6$ \\[1mm]
\hline
diffusivity & $D_r=D_z= 1e-3$ m$^2$/s \\[1mm]
\hline
friction & $\eta  = 1e5$ Pa s\\[1mm]
\hline
initial radius & $R_o=1500\,\mu$m\\[1mm]
\hline
initial swollen volume  \& stretch ratio & 
   $J_o=1000,\,\lambda_o=10$ \\[1mm]
\hline
initial aspect ratio & $AR=2\,R_o/H_o=20 \sim 40$ \\[1mm]
\hline
initial thickness & $H_o=150\,\mu m \sim 75\,\mu$m\\[1mm]
\hline
final volume/initial volume & $J_{a1}=0.05$ \\[1mm]
\hline
control time for $\beta$ & $\tau_\beta=20$ s \\[1mm]
\hline
control time for $\mu$ & $\tau_\mu=100$ s\\[1mm]
 \hline
 \end{tabular}
\end{table} 
The values assigned to the initial thickness and aspect ratio have been inspired by \cite{Ideses:2018}, and the successive parametric analyses always consider values of $AR$ and $H_o$ not too far from those ones. Moreover, we considered a highly swollen  initial state of the gel, which motivated our choice for the Flory parameter $\chi$ and the shear modulus $G$. Finally, as it has been observed in \cite{Ideses:2018} that the characteristic time of the process is about $200$s, and the characteristic time of the discharge velocity is about $40$s, we tuned the values assigned to the diffusion constants and to the friction in such a way to qualitative match the characteristic times of the dynamics.\\
Then, we fix the material and geometrical parameters as in Table \eqref{tab:para}. \\ 
\subsubsection{Initial state}
We assume a fully swollen state as initial state of the gel, characterized by an unstretched mesh size $\xi_a$ equal to the reference mesh size $\xi_d$. From an experimental point of view, it means that self contraction and liquid release are going to be initiated; from the modeling point of view, it means that 
\begin{equation}
\mu_{eo}=\mu_o=0\,\,\textrm{J/mol}\quad\textrm{and}\quad \quad J_{ao}=1\,.
\end{equation}
Given these values, we can use equations \eqref{ssfs2} to get the initial swollen state $J_o$ and 
the value of the active stress $\beta_o$ which maintains it: from 
%
\begin{equation}\label{is10}
0=\mu(J/J_{ao})\quad\textrm{and}\quad
 \beta_o=e_y(J_{ao},J_{eo})\,,
\end{equation}
we get $J_o$ and $\beta_o$. In particular, being $J_{eo}=J_o = \lambda_o^3$,
equation \eqref{is10}$_1$ takes the form 
\begin{equation}\label{is1}
0=    \left[ \log\left(\frac{\lambda_o^3-1}{\lambda_o^3}\right) + \frac{\chi+\lambda_o^3}{\lambda_o^6} \right] +\dfrac{m}{\lambda_o}\,,
\end{equation}
and can be solved for $\lambda_o$, the free-swelling stretch ratio at $\Bc_o$. Therein,
$m=G\Omega/R\,T$ is the ratio between the 
elastic energy and the mixing energy. 
Equation \eqref{is10}$_2$
determines the active stress $\beta_o$ (J/m$^3$) corresponding to null self-contraction ($\xi_a=\xi_d$) and to the free swelling stretch $\lambda_o$:
\begin{equation}\label{is2}
   \dfrac{\Omega}{R\,T}\beta_o=(\lambda_o^3-1)\left(\dfrac{\lambda_o^3-1}{\lambda_o^6}\chi -\dfrac{1}{\lambda_o^3}\right) + m\left(\dfrac{1}{\lambda_o}+\frac{\lambda_o^2}{2} -\frac{3}{2}\right)\,.
\end{equation}
It is worth noting that equation  \eqref{is1} is quite standard in stress-diffusion theories based on a Flory-Rehner thermodynamics \cite{FR1,FR2}; it is easy to verify that, given $\mu_o$,  the free-swelling stretch $\lambda_o$ increases as $m$ decreases, as shown in figure \ref{fig:1} (panel b), where the relation between $\lambda_o$ and $m$ has been represented. 
On the contrary, equation \eqref{is2} does not belong to standard stress-diffusion theory, and is peculiar of the present augmented model.  Figure \ref{fig:1} (panel b) also shows the dependence of $\beta_o$ on $\lambda_o$.\\
Finally, equations \eqref{is1} and \eqref{is2} deliver the following initial values of $J$, $J_e$, $c_d$, $p$ and $\beta$:
\begin{equation}
\begin{array}{l}
J_{eo}=J_o=1000\,, \\[2mm]
c_{do}=(J_o-1)/\Omega=5.5e7\,\,\textrm{mol/m$^3$}, \\[2mm]
p_o=G\,(1/J_{eo})^{1/3}= 13.6\,\,\textrm{Pa},  \\[2mm]
\beta_o=e_y(1,J_{eo})= -8e7\,\,\textrm{J/m$^3$}.\\[2mm]
\end{array}
\end{equation}
\subsubsection{Final states}
In the experiments, it has been observed the attainment of final steady states, when self contraction and liquid transport stop. In the modeling, we studied the conditions to get final steady states, which are not too far, in terms of some characteristic elements, from those experimental final states.\\
We considered two different protocols: (a), where  only active stresses $\beta\not=\beta_o$ drive the active contractions and liquid transport; (b),  where  bot active stresses $\beta\not=\beta_o$ and a change in the chemical potential of the bath from $\mu_o$ to $\mu_e$ drive the active contractions and liquid transport.
In both the protocols, based on the outcomes of the experiments presented in \cite{Ideses:2018}, we assume that the unstretched mesh size
is contracted by $\xi_a/\xi_d=J_{a1}^{1/3}\simeq 0.38$ with respect to 
the dry mesh size, and set the final value $J_{a1}$ of $J_a$ in such a way to produce that result: $J_{a1}=(\xi_a/\xi_d)^3=0.05$.\\
\paragraph{Protocol a} 
Assuming that
\begin{equation}
\mu_{e1}=\mu_o=\mu_1=0\,\textrm{J/mol}\,,
\quad J_{a1}=0.05\,,
\end{equation}
equations \eqref{ssfs2} deliver the final swelling ratio $J_1$ and 
the active stress $\beta_1$ needed to maintain it: 
%
\begin{equation}\label{is1A}
\mu(J/J_{a1})=0 ,\,
 e_y(J_{a1},J_{e1})=\beta_1
 \quad\Rightarrow\quad
 J_1,\,\,\beta_1
\end{equation}
Given our parameters, for case (a) we have the following characteristic values of the final state:
\begin{equation}
\begin{array}{l}
J_{e1}=J_1/J_{a1}=J_{eo}=1000\,,\quad J_1=50\,, \\[2mm]
c_{d1}=(J_1-J_{a1})/\Omega=2.8e7\,\,\textrm{mol/m$^3$}, \\[2mm]
\beta_1=e_y(J_{a1},J_{e1})= -4e6\,\,\textrm{J/m$^3$}.\\[2mm]
\end{array}
\end{equation}
It is worth noting that at the initial state, in absence of contraction ($J_{a1}=1$), we get $J_1=J_{e1}=1000$, whereas at the final contracted state $\Bc_1$, we get $J_1=J_{e1}=50$, that is,  a much smaller volume change under the same chemical conditions. It means that  the model includes an effective bulk stiffening of the gel due to self-contraction, that is, to motor activity, that has already been recognized as crucial in other works \cite{MacKintosh:2008}.\\
\paragraph{Protocol b} Typically, in a Lab, the chemical potential of the bath is not controlled. We can suppose it is constant, as in the protocol a) or, as it is  possible that some 
chains of the gel, which are not perfectly cross-linked, are released during the gel contraction, we can assume that it varies \cite{Haviv:2008}. This motivated our choice to follow protocol b), too. We assume that the final swelling ratio $J_1$ is half the value of case (a), 
while $J_{a1}$ is the same as before:
\begin{equation}
J_1=25\,,
\quad J_{a1}=0.05\,.
\end{equation}
Now, equations \eqref{ssfs2} are used to identify the final chemical potential $\mu_1$ and 
the active stress $\beta_1$ needed to maintain this final state: 
\begin{equation}\label{is1B}
\mu(J_1/J_{a1})=\mu_1 ,\,
 e_y(J_{a1},J_{e1})=\beta_1
 \,\Rightarrow\,
 \mu_1,\,\,\beta_1\,.
\end{equation}
Given the parameters, for case (b) we have the following characteristic values of the final state:
\begin{equation}
\begin{array}{l}
J_{e1}=J_1/J_{a1}=J_{eo}=500, \\[2mm]
p_1=G\,(J_{a1}/J_{1})^{1/3} = 17.1\,\,\textrm{Pa},  \\[2mm]
c_{d1}=(J_1-J_{a1})/\Omega=1.4e6\,\,\textrm{mol/m$^3$}, \\[2mm]
\mu_1=\mu(J_{a1},J_{e1})=-6.7e-4\,\,\textrm{J/mol}, \\[2mm]
\beta_1=e_y(J_{a1},J_{e1})= -4e6\,\,\textrm{J/m$^3$}.\\[2mm]
\end{array}
\end{equation}
We note that for the two cases, the value of $\beta_1$ is the same, but the
de-swollen volume $J_1$ is quite different ($50$ vs $25$), as for case (b)  liquid transport and release is driven by  both the mesh contraction and the change in the chemical conditions of the bath, whereas for the case a) only the driving force is only the gel activity.\\
%


%
\begin{figure}[h]
\centering
  \includegraphics[width=0.9\columnwidth]{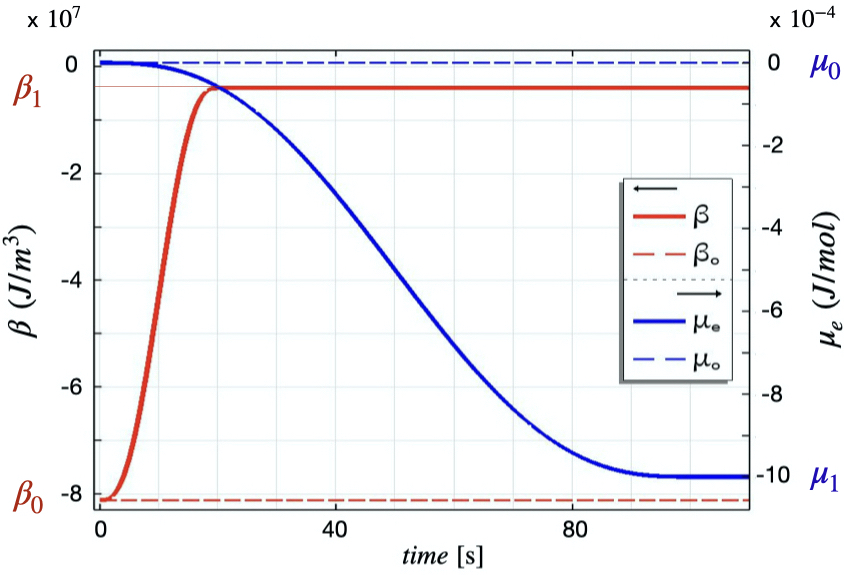}
  \caption{Time laws of the controls $\mu_e$ and $\beta$;
  for both cases (a) and (b), $\beta$ runs from  $\beta_o=-8\cdot 10^7$ J/m$^3$ (at time $\tau=0$ s) to $\beta_1=-.4\cdot 10^7$ J/m$^3$ (at time $\tau=\tau_\beta=20$ s) (dashed \& solid red).
  Case a): $\mu=\mu_o=0$ J/mol  (at time $\tau=0$ s) (dashed blue).
  Case b): $\mu$ runs from  $\mu_o=0$ J/mol to $\mu_1=-10^{-3}$ J/mol  (at time $\tau=\tau_\mu=100$ s) (solid blue). $\beta$ axis at left, $\mu$ axis at right.}
  \label{fig:3}
\end{figure}
\begin{figure}[h]
\centering
  \includegraphics[width=0.8\columnwidth]{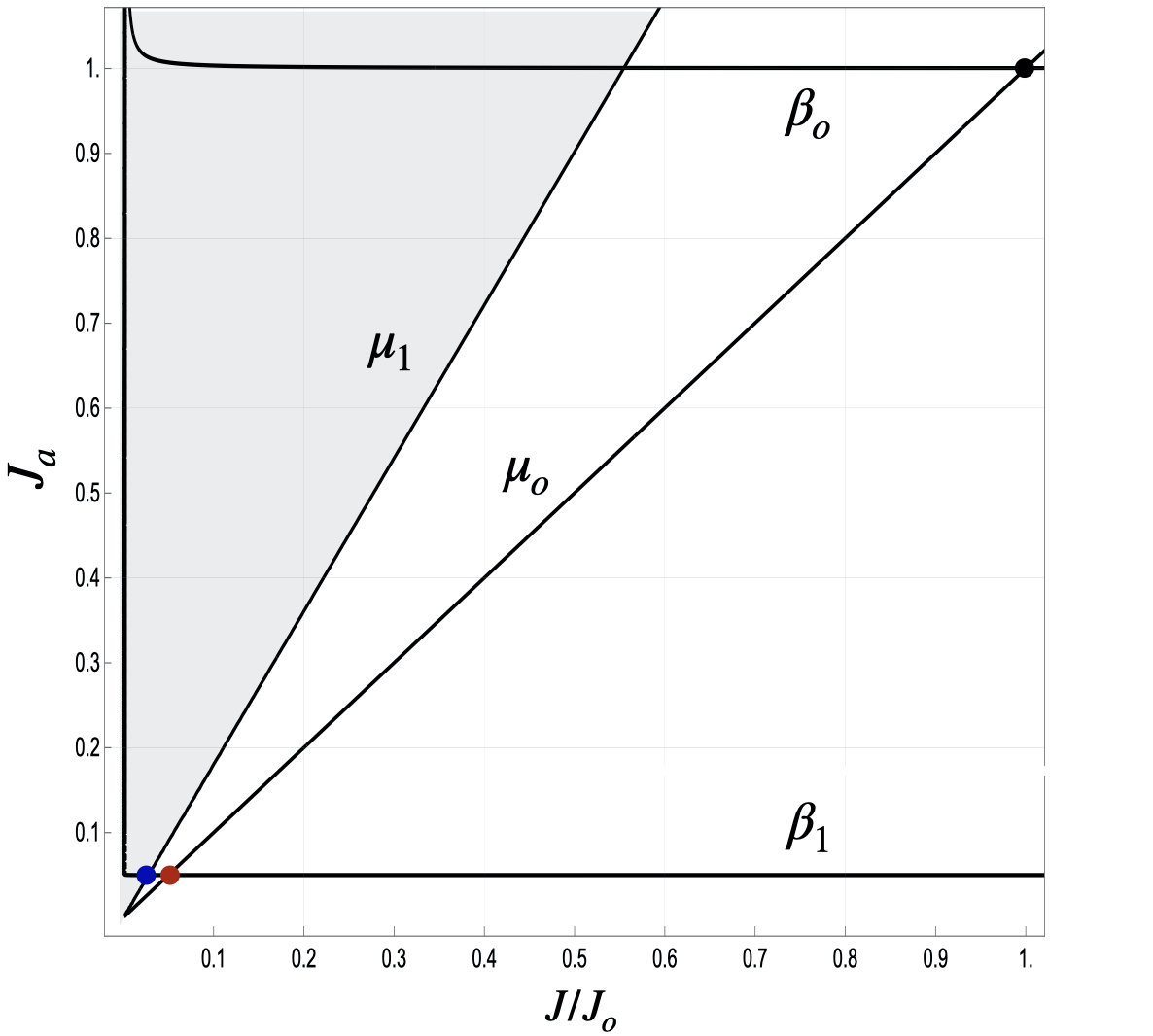}
  \caption{Swelling-contraction diagram $J_a$ versus $J/J_o$ at equilibrium and stress-free states. The isolines $\mu=\mu_o$ and $\mu=\mu_1$ are identified by straight lines in this diagram, where the isoline $\beta=\beta_o$ and $\beta=\beta_1$ are hyperboles. The black dot corresponds to the initial state whereas the red and blue dots corresponds to the final state attained under case a) and b), respectively.}
  \label{fig:4}
\end{figure}
%
%
%
\section{Contraction dynamics} \label{SSS}
%
Our idea is that geometry greatly affect contraction dynamics due to the liquid transport, which has its own characteristic length, diversely from self-contraction dynamics, which don't have it since motor activity is  homogenous across the system.\\
 The key geometrical parameter in a disc is its  aspect ratio $AR$; hence, we start investigating the effects of $AR$ on the contraction dynamics with two complementary studies: 1) at fixed radius $R_o=1.5$ mm, and 
varying $H_o$;  2) at fixed thickness $H_o=0.10$ mm and varying $R_o$. The investigated range of parameter $AR$ is described in Table \ref{tab:AR}: the analysis goes from discs whose AR varies from $20$ (thick discs) to discs of aspect ratio $45$ (thin discs). 
The study is carried on under the conditions of scenario a).
\begin{table}[h]
\caption{\label{tab:2} Data about aspect ratios; values of $R_o$ and $H_o$ are in mm \label{tab:AR}}
\centering
\begin{tabular}{lll}
\hline
AR	& $R_o (H_o=0.1)$ & $H_o (R_o=1.5$)\\[1mm]
\hline
$20$	& $1.0$ & $0.15$ \\[1mm]
\hline
$25$	& $1.25$ & $0.12$\\[1mm]
\hline
$30$     & $1.50$ & $0.1$ \\[1mm]
\hline
$35$  & $1.75$ & $0.086$ \\[1mm]
\hline
$40$ & $2.0$ & $0.075$ \\[1mm]
\hline
$45$ & $2.25$ & $0.066$\\[1mm]
\hline
\end{tabular}
\end{table} 
\\
All the experiments start with $J_o=1000$, a highly swollen initial state, and $J_{ao}=1$, and evolve towards the new steady values $J_1=50$ and $J_{a1}=0.05$. 
As stated above, these values correspond to a reduction in mesh = $\xi_{a1}/\xi_{ao} = 0.05^{1/3} = 0.38$, where $\xi_{a1}$ represents the final mesh size at zero stress, see Section \ref{IFS}.

In the regime under study, the system reaches the new equilibrium state at a time $\tau_1\simeq 200$ s, that is, we have  $\tau_\beta<<\tau_1$ and  dynamics is ruled by the redistribution of water, which has a length scale that is  the disc thickness $H_o$.\\
To present our results, we focus on: evolution
paths in the plane $(\bar{J},\bar{J}_a)$; 
velocities of the lateral boundary of the disc, \textit{i.e.}, radial velocity; radius and thickness reduction. The averages $\bar{J}$ and $\bar{J}_a$ of the fields $J(r, z, \tau)$ and $J_a(r, z, \tau)$ are introduced to give a global glance at the contraction dynamics and, due to the cylindrical symmetry of the system, are averaged on the cross section $\Sc_d$ of area $R_d\cdot H_d$. Changes in volume, boundary velocities and changes in radius and thickness have a large effect on the global change in shape of the disc. They are visible  in experiments and can be measured, if the appropriate tests are performed. Finally, we analyse the stress state of the gel during the contraction process.
\subsection{Dynamics in the plane $(\bar{J},\bar{J}_a)$}
\begin{figure}[h]
\centering
  \includegraphics[width=0.95\columnwidth]
  {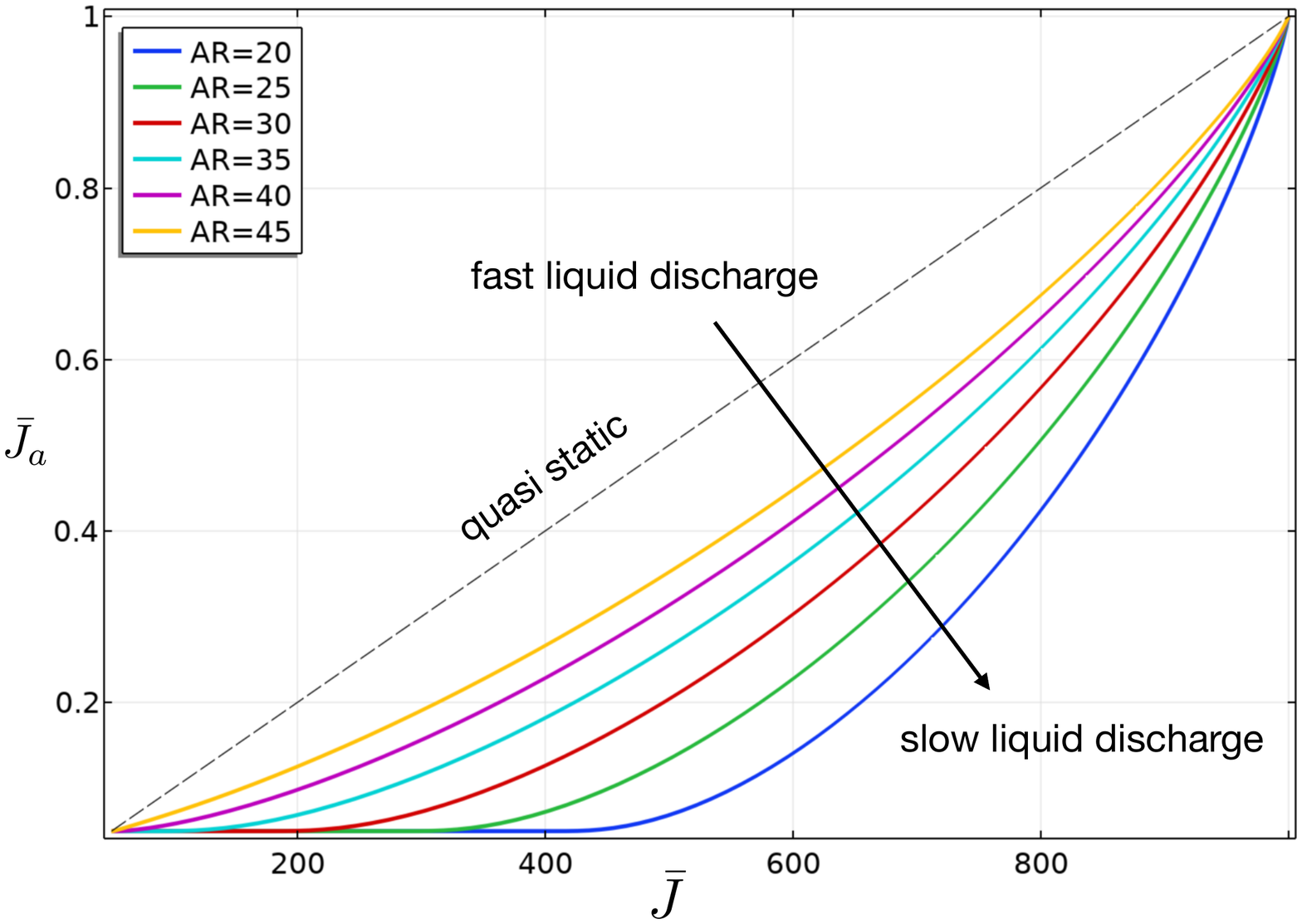}
  \caption{Plane $(\bar{J},\bar{J}_a)$: evolution path at constant radius $R_o=1.5$mm for different values of the aspect ratio AR.  Lower AR correspond to evolution path far from equilibrium; higher AR correspond to paths which tend to the quasi-static stress-free path (dashed line).}
  \label{fig:5}
\end{figure}

\begin{figure}[h]
\centering
  \includegraphics[width=0.95\columnwidth]{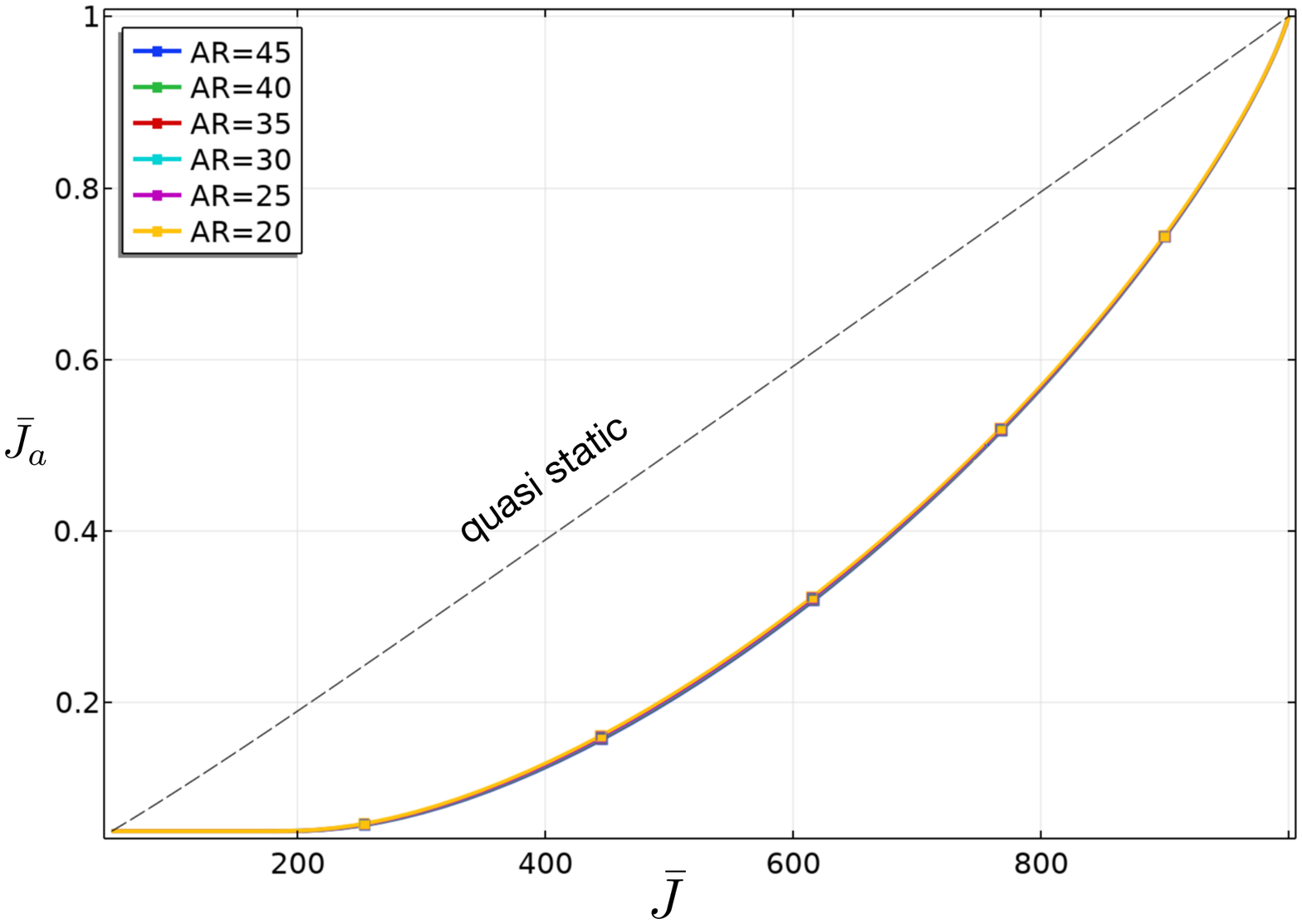}
  \caption{Plane $(\bar{J},\bar{J}_a)$: evolution path at constant thickness $H_o=0.1$mm for different values of the aspect ratio AR.  All the paths are superimposed and the master curve is the one corresponding to $AR=30$ in figure \ref{fig:5}.}
  \label{fig:6}
\end{figure}

The main features of the contraction dynamics are 
well represented  by the curves $\tau\mapsto(\bar{J}(\tau),\bar{J}_a(\tau))$, which are plotted
in the plane  $(\bar{J},\bar{J}_a)$. That plane allows us to glance at the quasi-static stress-free path characteristic of  an evolution which occurs as  a  sequence of equilibrium states (straight dashed line).  Thinner discs (higher AR) show an evolution in the plane which is closer to the stress-free path. Under the same contraction dynamics, liquid transport is faster for those discs and it allows to quickly recover the original stress-free state. On the contrary, for ticker discs (lower AR) the evolution path  is very far from the quasi-static regime: namely, motor-induced contraction is faster than the water trasnport across the gel pores, which makes the gel highly stressed during its evolution.\\
Figures \ref{fig:5} and \ref{fig:6}
show evolution paths for
different $AR$ for varying $H_o$ at constant $R_o$ (figure \ref{fig:5}) and varying $R_o$ at constant $H_o$ (figure \ref{fig:6}). In the first case, it is shown as decreasing the thickness $H_o$, that is, the characteristic length scale across which water flows, decreases the characteristic time scale of water trasnport (from blue to yellow solid lines).  Interestingly, 
in the second case, that is, changing AR by varying radius under constant thickness, we get a series of fully overlapped curves,  so confirming that  the important length scale for water exit is $H_o$.
%
%
\subsection{Gel contraction velocity}
\begin{figure}[h]
\centering
\includegraphics[width=0.95\columnwidth]{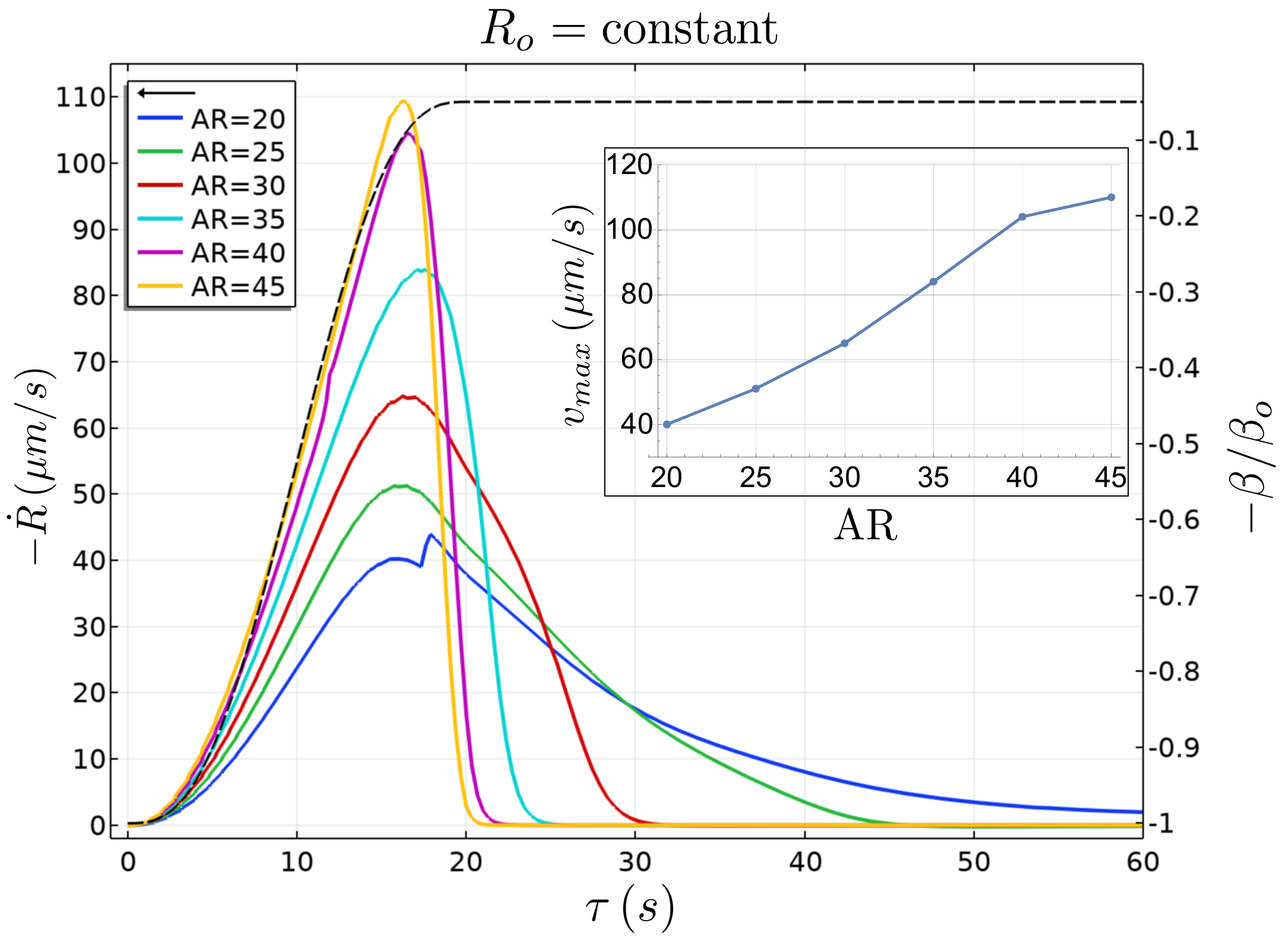}
\caption{Radial contraction velocity $\dot R$ of the lateral boundary of the disc, for different values of $AR$ at constant radius (solid); time evolution of $\beta$ (dashed). The color code is the same as in \ref{fig:5} and \ref{fig:6}. The small wiggle in the blue line at
$\tau\simeq 17$ s is due to a mechanical buckling: the disc departs from the flat shape,
see figure \ref{fig:10}, panel (C). 
Velocity ranges over  left vertical axis and $\beta/\beta_o$ over right vertical axis.}
\label{fig:7}
\end{figure}

\begin{figure}[h]
\centering
\includegraphics[width=0.95\columnwidth]{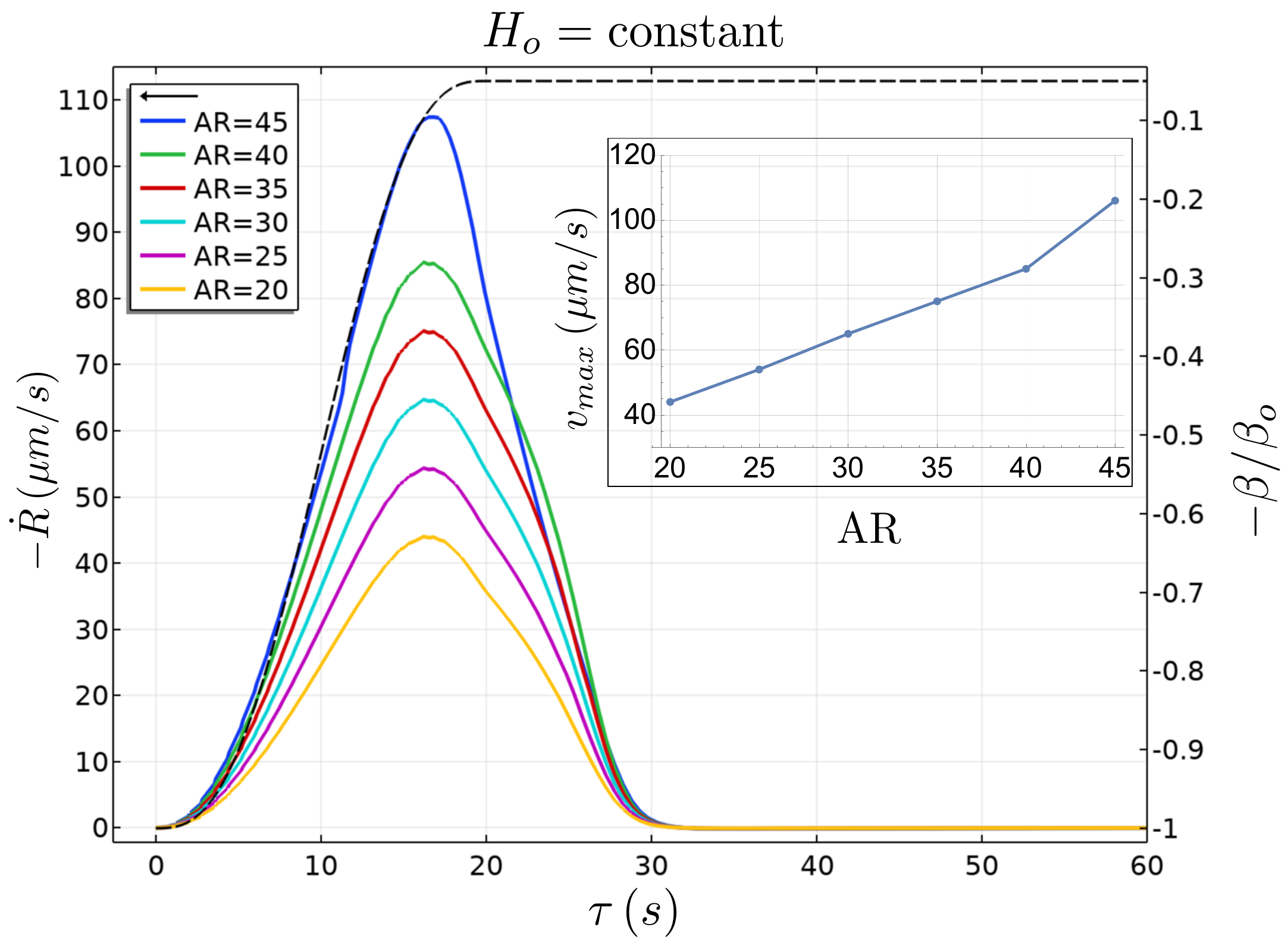}
\caption{Radial contraction velocity $\dot R$ of the lateral boundary of the disc, for different values of $AR$ at
constant thickness (solid) and evolution of $\beta$ (dashed). 
Color code is the same as in figures \ref{fig:5} and \ref{fig:6}.
Velocity ranges over  left vertical axis and $\beta/\beta_o$ over right vertical axis.}
\label{fig:8}
\end{figure}

%
Through the aforementioned studies, we investigate also the effects of $AR$ on the radial contraction velocity $\dot{R}(\tau)$ of the lateral boundary of the gel disc.\\
The radial contraction velocity $\dot{R}(\tau)$ is determined from the  average current radius $R(\tau)$
\begin{equation}\label{Rt}
R(\tau)=\Lambda_r(\tau)R_d\quad\textrm{such that}\quad
\dot R(\tau) = \dot\Lambda_r(\tau)\,R_d\,,
\end{equation}
where $\Lambda_r$  is an average stretch defined as
\begin{equation}\label{Lrt}
\Lambda_r(\tau)=1 + \frac{1}{H_d}\int_0^{H_d}\frac{u(R_d,z,\tau)}{R_d}\,dz\,.
\end{equation}
From \eqref{Rt} and \eqref{Lrt}, the radial contraction velocity can be also rewritten as $\dot R(\tau)=\dot\Lambda_r(\tau)\,\dfrac{H_d}{2}\,AR$. It is easy to verify  that the average stretch $\Lambda_r$ also corresponds to the average $\bar\lambda_r(\tau)$ of the radial deformation $\lambda_r(r,z,\tau)$  on the cross section $\Sc_d$ of area $R_d\cdot H_d$.\\
The numerical results obtained for a constant radius  show that the radial velocity $\dot R(\tau)$  is characterized by two time scales, one that characterizes the phase in which the velocity increases and the second of the velocity decrease phase  (figure \ref{fig:7}). 
During the growth phase, the curves fit to a linear law, that is, 
$\dot R(\tau)\propto \tau/\tau_r$ with $\tau_r$ the
characteristic time of rising. During the decreasing phase, curves fit to an exponential law 
$\dot R(\tau)\propto v_{max}\exp(-\,\tau/\tau_{decay})$, with $\tau_{decay}$ the characteristic time of decay (see Table \ref{tab:2}).\\
\begin{table}[h]
\caption{\label{tab:2} Max velocity, peak time, rising time and decay time, for different values of aspect ratio AR}
\centering
\begin{tabular}{lllll}
\hline
AR & $v_{max} \mu$ m/s & $\tau_p$  & $\tau_r$  & $\tau_{\rm{decay}}$  \\[1mm]
\hline
20 & $44 \mu$m/s & $ \simeq 16$ s & $0.22$ s & $13$ s \\[1mm]
\hline
25 & $52 \mu$m/s & $ \simeq 16$ s & $0.22$ s & $13$ s \\[1mm]
\hline
30 & $74 \mu$m/s & $ \simeq 17$ s & $0.18$ s & $8$ s \\[1mm]
\hline
35 & $84 \mu$m/s & $ \simeq 17$ s & $0.14$ s & $3$ s \\[1mm]
\hline
40 & $104 \mu$m/s & $ \simeq 17$ s & $0.12$ s & $2$ s \\[1mm]
\hline
45 & $111 \mu$m/s & $ \simeq 17$ s & $0.11$ s & $1.5$ s \\[1mm]
\hline
\end{tabular}
\end{table} 
The inset in the figure shows that the maximum radial velocity $v_{max}$, attained  at peak time $\tau_p$, depends
on the geometric parameter $R_o$ and $H_o$.\\ 
Actually, the analysis of the equations \eqref{Rt} and \eqref{Lrt} shows that when $AR$ changes with $H_d$ (or, equivalently, with $H_o$ as the initial free-swelling is homogeneous), with $R_o$ constant, the dependence of $\dot R$ on $AR$ is also affected by $H_d$  and can't be linear. The same equations show that, for $H_d$ constant the dependence of $\dot R$ on $AR$ is simply linear. This is what the inset in figure \ref{fig:7} shows for the maximum velocity $v_{max}$, relative to the study at varying radius.\\
Moreover, 
we can split the average stretch $\Lambda_r$ into an elastic component $\Lambda_e$ and an active component $\Lambda_a$, related to the analogous multiplicative decomposition of the deformation gradient $\Fb=\Fb_e\Fb_a$ and of the radial deformation $\lambda_r$. Thus, the stretching velocity $\dot\Lambda_r$ can be additively split in two summands:
\begin{equation}\label{udot2}
\dot R =
(\dot\Lambda_a\,\Lambda_e + \Lambda_a\,\dot\Lambda_e )\,R_d\,, 
 \end{equation}
where $\Lambda_a$ and $\Lambda_e$ are defined as the average of the active $\gamma_r$ and elastic $\lambda_r/\gamma_r$ radial deformation, with first due to self-contraction and the second driven by liquid transport. 
Equation \eqref{udot2} highlights the existence
of two time scales for $\dot R$:
for $\tau\le\tau_\beta$ the stretching velocity is dominated by the time evolution of $\beta(\tau)$, while for 
$\tau\ge\tau_\beta$ is dominated by solvent release; we have
\begin{equation}\label{udot3}
\begin{array}{lcll}
\dot R &\simeq&
\dot\Lambda_a\,\Lambda_e \,R_d 
 & \textrm{$\tau < \tau_\beta$, contraction-dominated regime}\,,\\[2mm]
\dot R &\simeq & \Lambda_a\,\dot\Lambda_e\,R_d &\textrm{$\tau > \tau_\beta$, liquid-dominated regime}\,.
\end{array}
\end{equation}
Equation \eqref{udot3}$_1$ shows that during the contraction-dominated regime, that is, for $t<\tau_\beta=20$ s,  the radial velocity $\dot R$ changes with the same rate of $\Lambda_a$, which depends on $\beta$, as figures \ref{fig:7} and \ref{fig:8} show (compare the coloured lines with the dashed black line in both figures).
On the other side, equation \eqref{udot3}$_2$ shows that during the liquid-dominated regime, that is, for $t>\tau_\beta=20$ s,  the radial velocity $\dot R$ changes with the rate of $\Lambda_e$, which depends on liquid transport and on the $AR$ of the disc, as figure \ref{fig:7} shows.\\
Figures \ref{fig:7} and \ref{fig:8} show also clearly that the maximal velocity is reached when contraction is maximal - as  was suggested in \cite{Ideses:2018} (see figure 4f in \cite{Ideses:2018}).\\

It is worth noting that the remodeling action $\beta$, needed to change the  target mesh size, does not further change once it has taken its maximal value. Beyond that,  the system evolves towards its steady state by releasing liquid and the steady state is reached when motor applied activity stresses are balanced by network elasticity such that the system reaches a stress free configuration.\\ 
It is also worth noting that the difference in the  behaviour of the $\dot R$ vs time curves  in the liquid-dominated regime for $R_o$ (figure \ref{fig:7}) and $H_o$ (figure \ref{fig:8}) constant is different as for these geometries thickness is important. 
%
%

%
\subsection{Stress distribution}
%
\begin{figure}[h]
\centering
\includegraphics[width=0.95\columnwidth]{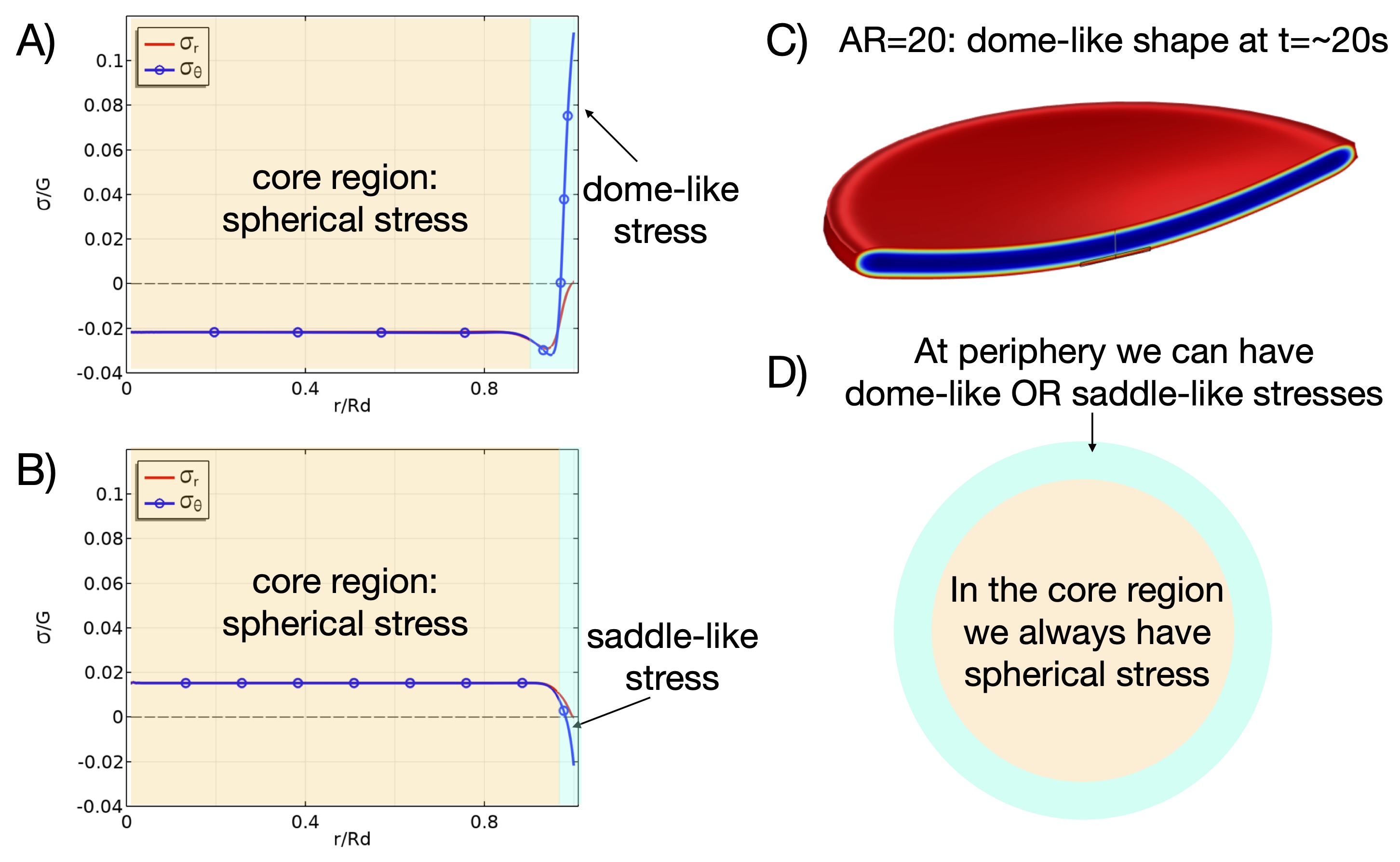}
\caption{Effect of $AR$ on stress distribution for simulations at constant radius.
Panels A) and B) show the radial $\sigma_r$ (red) and hoop $\sigma_\theta$ (blue) stresses versus the non dimensional radius $r/R_d$ at $\tau=20$ s, for $AR=20$ and $AR=45$. 
A) $AR=20$: the hoop stress is negative in the core 
(beige) and positive at the periphery (cyan), a typical pattern of frustrated dome-like shape.
\label{fig:10}}

%
\end{figure}
Stress analysis in the active disc can be relevant, as it might drive mechanical instability, which lead to a variety of different shapes at the end of the contraction\cite{Efrati:2009,Sharon:2010,Pezzulla:2015,Ideses:2018}.
The analysis of instabilities is beyond the scope of the present work, and will mark our future efforts. However, through the aforementioned studies, we might have interesting clues about shape transitions by 
investigating the effects of $AR$ on the
the evolution of radial and hoop stresses in the disc, which may drive further experiments.

We only report results for the case of constant radius. We compare the stress state in a  thick ($AR\simeq 20$) and a thin ($AR\simeq 45$) disc. 
Panels A) and B) of figure \ref{fig:10} show the existence of two stress patterns: stress is constant in a core region (beige) and varying in a peripheral one (cyan). 
As bulk contraction $\beta$ is homogeneous and isotropic in the whole disc, these two regions are determined by the dynamics of liquid transport.
In particular, the width of the peripheral region is of the order of the thickness because the solvent in this region can escape from both the lateral boundary and the top and bottom surfaces. In contrast, for the solvent in the core the shortest path to exit the gel disc is through the top and bottom surfaces.\\
In particular, in figure \ref{fig:10}, for $AR=20$ we have essentially $\sigma_r < 0$ along all the radius (see panel A)
, and $\sigma_\theta$ varying from negative to positive,
(see panel A); for $AR=45$ we have $\sigma_r > 0$ along all the radius (see panel B)
and $\sigma_\theta$ varying from positive to negative (see panel B).
Corresponding to our values of $AR$, we have $H_{\rm{thin}}\simeq 0.04\,R_d$
and $H_{\rm{thick}}= 0.1\,R_d$.
The stress distribution for the two cases is typical of
that found in frustrated dome-like or saddle-like discs (see figure \eqref{fig:10}, panels C and D)\cite{Efrati:2009,Sharon:2010,Pezzulla:2015}.

That is a preliminary requirement for observing instability patterns which can deliver domes or saddles, depending on other key factors, which are not investigated in the present paper.
%
\subsection{Evolution of aspect ratio during contraction}
%
Finally, the geometry of the gel body suggested to investigate the possibility to have frictions $\eta_r$ and $\eta_\theta$ in the plane, different from the vertical friction $\eta_z$. Frictions are related to the resistances of the mesh to remodel, which can be expected to be different. Our conjecture needs to be validated and the analysis may stimulate further experiments in this direction.\\
As noted at the end of Section \ref{DIC}, the
system is controlled by the pair $(\mu_e,\beta)$, and here we also analyse the combined effects of varying the chemical potential
$\mu_e$ and active force $\beta$ (protocol b).\\
We model the motor activity by introducing a uniform and isotropic active stress $\beta$.
Nevertheless, during gel contraction, the  radial and vertical stretches might differ locally and each one of them can vary in time and space. We use the average values $R(\tau)$ and $H(\tau)$, defined as $H(\tau) = \Lambda_z(\tau)\,H_d$ with
\begin{equation}
\Lambda_z(\tau)=1 + \frac{1}{R_d}\int_0^{R_d}\frac{w(r,H_d,\tau)}{H_d}\,dr\,, 
\end{equation}
to describe the change in the aspect ratio of the disc. 
\begin{figure}[h]
\centering
  \includegraphics[width=0.95\columnwidth]{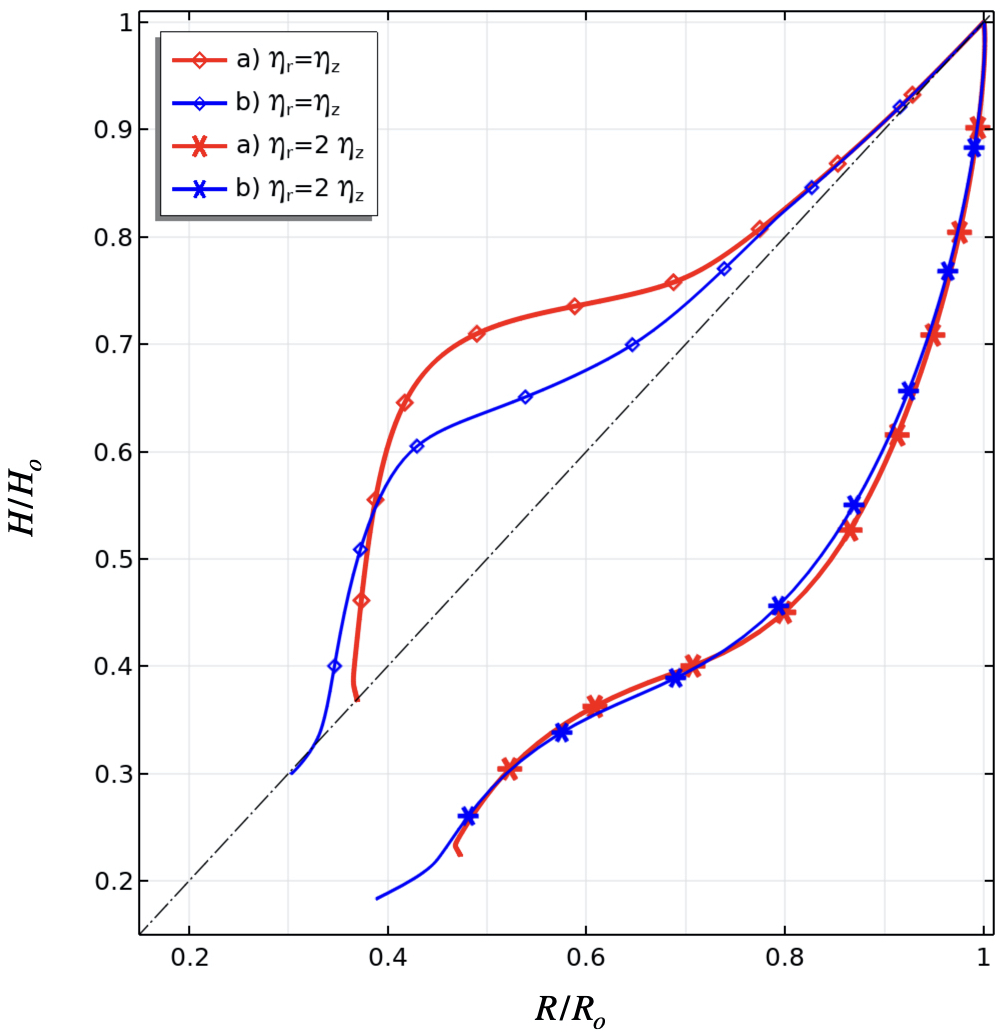}  \caption{Thickness ratio $H/H_o$ versus radius ratio $R/R_o$ during contraction for cases a) (red) and b) (blue) with equal friction $\eta_r=\eta_z$ (diamond) and differential friction $\eta_r=2\,\eta_z$ (star); $\eta_r=10^5$ Pa s. Dashed line represents isotropic contractions;  with different frictions, the radial and vertical contractions are not isotropic. Disc geometry: $R_o=1.5$ mm, $AR=22$.}
  \label{fig:11}
\end{figure}
At any time $\tau$, the ratio ${H}(\tau)/H_o$ can be plotted against the ratio ${R}(\tau)/R_o$ to illustrate the evolution path of the radial and vertical stretches, that is
the curve $\tau\mapsto({R}(\tau)/R_o, {H}(\tau)/H_o)$, plotted
in the plane $({R}/R_o,{H}/H_o)$. In  figure \ref{fig:11}), the curve has been represented for a disc with $AR=22$ and $R_o=1.5$ mm. In that plot, the dashed line represents
an isotropic evolution, during which the aspect ratio
remains constant during network contraction.\\
For each of the two analyzed cases a) (red) and b) (blue), we show two curves, one corresponding to equal frictions (diamond),
$\eta_r=\eta_\theta=\eta_z$,
and the other with different horizontal and vertical frictions (asterisk),
$\eta_r=\eta_\theta=2\,\eta_z$.
We note that the evolution is very sensitive to friction, 
while the differences between case a) and b) are 
less noticeable. For all simulations, the system evolves via a characteristic path. It departs from the isotropic contraction
path, but in the case with equal frictions the steady state configuration ends on the dashed line (i.e., on the isotropic path), while the case with different frictions ends far from it.
In particular, when $\eta_r=\eta_z$, the contraction is almost isotropic until $H/H_o=R/R_o\sim 0.8$; then, radial contraction is faster, and eventually the vertical one becomes faster.
When $\eta_r=2\,\eta_z$,  vertical contraction is much faster than the radial one, and the final state is not isotropic.\\
%
%
\section{Conclusions and future directions}
We discussed the interplay between elasticity, liquid transport and self-contractions in active gel discs from the perspective of continuum mechanics. It has been shown that, even if  contraction dynamics  doesn't have a  characteristic length, the aspect ratio of active gel discs may greatly affect the changes in shape, due to the  dependence of contraction dynamics on liquid transport, which is system-size dependent.\\
To keep the model easy, the numerical model has been developed under the hypothesis of cylindrical symmetry, which excludes the challenge to observe disc morphings  which are not compatible with the cylindrical symmetry.
Actually, we are planning to give up the symmetry hypothesis above and investigate the blossom of stresses in the disc, which may drive instability patterns and, consequently, a variety of steady shapes of the gel. It was  beyond the scope of the present work and it'll mark our future efforts.\\
Giving up the symmetry hypothesis makes also more interesting  the identification of the determinants of possible  changes in shape, whose control  would open to the possibility to get actuators based on self-contractile gels, a promising field which can be set within the framework here presented.
%
%

\begin{acknowledgements}
This work has been supported by MAECI (Ministry of Foreign Affairs and International Cooperation) and MOST (Ministry of Science and Technology - State of Israel) through the project PAMM. F.R. also thanks INDAM-GNFM for being supported with Progetti Giovani GNFM 2020.
\end{acknowledgements}

\bibliography{2022_gel}

\begin{thebibliography}{31}%
\makeatletter
\providecommand \@ifxundefined [1]{%
 \@ifx{#1\undefined}
}%
\providecommand \@ifnum [1]{%
 \ifnum #1\expandafter \@firstoftwo
 \else \expandafter \@secondoftwo
 \fi
}%
\providecommand \@ifx [1]{%
 \ifx #1\expandafter \@firstoftwo
 \else \expandafter \@secondoftwo
 \fi
}%
\providecommand \natexlab [1]{#1}%
\providecommand \enquote  [1]{``#1''}%
\providecommand \bibnamefont  [1]{#1}%
\providecommand \bibfnamefont [1]{#1}%
\providecommand \citenamefont [1]{#1}%
\providecommand \href@noop [0]{\@secondoftwo}%
\providecommand \href [0]{\begingroup \@sanitize@url \@href}%
\providecommand \@href[1]{\@@startlink{#1}\@@href}%
\providecommand \@@href[1]{\endgroup#1\@@endlink}%
\providecommand \@sanitize@url [0]{\catcode `\\12\catcode `\$12\catcode
  `\&12\catcode `\#12\catcode `\^12\catcode `\_12\catcode `\%12\relax}%
\providecommand \@@startlink[1]{}%
\providecommand \@@endlink[0]{}%
\providecommand \url  [0]{\begingroup\@sanitize@url \@url }%
\providecommand \@url [1]{\endgroup\@href {#1}{\urlprefix }}%
\providecommand \urlprefix  [0]{URL }%
\providecommand \Eprint [0]{\href }%
\providecommand \doibase [0]{https://doi.org/}%
\providecommand \selectlanguage [0]{\@gobble}%
\providecommand \bibinfo  [0]{\@secondoftwo}%
\providecommand \bibfield  [0]{\@secondoftwo}%
\providecommand \translation [1]{[#1]}%
\providecommand \BibitemOpen [0]{}%
\providecommand \bibitemStop [0]{}%
\providecommand \bibitemNoStop [0]{.\EOS\space}%
\providecommand \EOS [0]{\spacefactor3000\relax}%
\providecommand \BibitemShut  [1]{\csname bibitem#1\endcsname}%
\let\auto@bib@innerbib\@empty
\bibitem [{\citenamefont {Bendix}\ \emph {et~al.}(2008)\citenamefont {Bendix},
  \citenamefont {Koenderink}, \citenamefont {Cuvelier}, \citenamefont {Dogic},
  \citenamefont {Koeleman}, \citenamefont {Brieher}, \citenamefont {Field},
  \citenamefont {Mahadevan},\ and\ \citenamefont {Weitz}}]{Bendix:2008}%
  \BibitemOpen
  \bibfield  {author} {\bibinfo {author} {\bibfnamefont {P.~M.}\ \bibnamefont
  {Bendix}}, \bibinfo {author} {\bibfnamefont {G.~H.}\ \bibnamefont
  {Koenderink}}, \bibinfo {author} {\bibfnamefont {D.}~\bibnamefont
  {Cuvelier}}, \bibinfo {author} {\bibfnamefont {Z.}~\bibnamefont {Dogic}},
  \bibinfo {author} {\bibfnamefont {B.~N.}\ \bibnamefont {Koeleman}}, \bibinfo
  {author} {\bibfnamefont {W.~M.}\ \bibnamefont {Brieher}}, \bibinfo {author}
  {\bibfnamefont {C.~M.}\ \bibnamefont {Field}}, \bibinfo {author}
  {\bibfnamefont {L.}~\bibnamefont {Mahadevan}},\ and\ \bibinfo {author}
  {\bibfnamefont {D.~A.}\ \bibnamefont {Weitz}},\ }\bibfield  {title} {\bibinfo
  {title} {{A Quantitative Analysis of Contractility in Active Cytoskeletal
  Protein Networks}},\ }\href {https://doi.org/10.1529/biophysj.107.117960}
  {\bibfield  {journal} {\bibinfo  {journal} {Biophysical Journal}\ }\textbf
  {\bibinfo {volume} {94}},\ \bibinfo {pages} {3126} (\bibinfo {year}
  {2008})}\BibitemShut {NoStop}%
\bibitem [{\citenamefont {Koenderink}\ \emph {et~al.}(2009)\citenamefont
  {Koenderink}, \citenamefont {Dogic}, \citenamefont {Nakamura}, \citenamefont
  {Bendix}, \citenamefont {MacKintosh}, \citenamefont {Hartwig}, \citenamefont
  {Stossel},\ and\ \citenamefont {Weitz}}]{Koenderink:2009}%
  \BibitemOpen
  \bibfield  {author} {\bibinfo {author} {\bibfnamefont {G.~H.}\ \bibnamefont
  {Koenderink}}, \bibinfo {author} {\bibfnamefont {Z.}~\bibnamefont {Dogic}},
  \bibinfo {author} {\bibfnamefont {F.}~\bibnamefont {Nakamura}}, \bibinfo
  {author} {\bibfnamefont {P.~M.}\ \bibnamefont {Bendix}}, \bibinfo {author}
  {\bibfnamefont {F.~C.}\ \bibnamefont {MacKintosh}}, \bibinfo {author}
  {\bibfnamefont {J.~H.}\ \bibnamefont {Hartwig}}, \bibinfo {author}
  {\bibfnamefont {T.~P.}\ \bibnamefont {Stossel}},\ and\ \bibinfo {author}
  {\bibfnamefont {D.~A.}\ \bibnamefont {Weitz}},\ }\bibfield  {title} {\bibinfo
  {title} {An active biopolymer network controlled by molecular motors},\
  }\href {https://doi.org/10.1073/pnas.0903974106} {\bibfield  {journal}
  {\bibinfo  {journal} {Proceedings of the National Academy of Sciences}\
  }\textbf {\bibinfo {volume} {106}},\ \bibinfo {pages} {15192} (\bibinfo
  {year} {2009})},\ \Eprint
  {https://arxiv.org/abs/https://www.pnas.org/content/106/36/15192.full.pdf}
  {https://www.pnas.org/content/106/36/15192.full.pdf} \BibitemShut {NoStop}%
\bibitem [{\citenamefont {Matthias~Schuppler}\ and\ \citenamefont
  {Bausch}(2016)}]{Schuppler:2016}%
  \BibitemOpen
  \bibfield  {author} {\bibinfo {author} {\bibfnamefont {M.~K.}\ \bibnamefont
  {Matthias~Schuppler}, \bibfnamefont {Felix C.~Keber}}\ and\ \bibinfo {author}
  {\bibfnamefont {A.~R.}\ \bibnamefont {Bausch}},\ }\bibfield  {title}
  {\bibinfo {title} {Boundaries steer the contraction of active gels},\
  }\href@noop {} {\bibfield  {journal} {\bibinfo  {journal} {Nature
  Communications}\ }\textbf {\bibinfo {volume} {7}},\ \bibinfo {pages} {13120}
  (\bibinfo {year} {2016})}\BibitemShut {NoStop}%
\bibitem [{\citenamefont {Bernheim-Groswasser}\ \emph {et~al.}()\citenamefont
  {Bernheim-Groswasser}, \citenamefont {Gov}, \citenamefont {Safran},\ and\
  \citenamefont {Tzlil}}]{Bernheim:2018}%
  \BibitemOpen
  \bibfield  {author} {\bibinfo {author} {\bibfnamefont {A.}~\bibnamefont
  {Bernheim-Groswasser}}, \bibinfo {author} {\bibfnamefont {N.~S.}\
  \bibnamefont {Gov}}, \bibinfo {author} {\bibfnamefont {S.~A.}\ \bibnamefont
  {Safran}},\ and\ \bibinfo {author} {\bibfnamefont {S.}~\bibnamefont
  {Tzlil}},\ }\bibfield  {title} {\bibinfo {title} {Living matter: Mesoscopic
  active materials},\ }\href {https://doi.org/10.1002/adma.201707028}
  {\bibfield  {journal} {\bibinfo  {journal} {Advanced Materials}\ }\textbf
  {\bibinfo {volume} {30}},\ \bibinfo {pages} {1707028}}\BibitemShut {NoStop}%
\bibitem [{\citenamefont {Ideses}\ \emph {et~al.}(2018)\citenamefont {Ideses},
  \citenamefont {Erukhimovitch}, \citenamefont {Brand}, \citenamefont
  {Jourdain}, \citenamefont {Salmeron}, \citenamefont {Gabinet}, \citenamefont
  {Safran}, \citenamefont {Kruse},\ and\ \citenamefont
  {Bernheim-Groswasser}}]{Ideses:2018}%
  \BibitemOpen
  \bibfield  {author} {\bibinfo {author} {\bibfnamefont {Y.}~\bibnamefont
  {Ideses}}, \bibinfo {author} {\bibfnamefont {V.}~\bibnamefont
  {Erukhimovitch}}, \bibinfo {author} {\bibfnamefont {R.}~\bibnamefont
  {Brand}}, \bibinfo {author} {\bibfnamefont {D.}~\bibnamefont {Jourdain}},
  \bibinfo {author} {\bibfnamefont {J.}~\bibnamefont {Salmeron}, \bibfnamefont
  {Hernandez}}, \bibinfo {author} {\bibfnamefont {U.}~\bibnamefont {Gabinet}},
  \bibinfo {author} {\bibfnamefont {S.}~\bibnamefont {Safran}}, \bibinfo
  {author} {\bibfnamefont {K.}~\bibnamefont {Kruse}},\ and\ \bibinfo {author}
  {\bibfnamefont {A.}~\bibnamefont {Bernheim-Groswasser}},\ }\bibfield  {title}
  {\bibinfo {title} {Spontaneous buckling of contractile poroelastic actomyosin
  sheets},\ }\href@noop {} {\bibfield  {journal} {\bibinfo  {journal} {Nature
  Communications}\ }\textbf {\bibinfo {volume} {9}},\ \bibinfo {pages} {2461}
  (\bibinfo {year} {2018})}\BibitemShut {NoStop}%
\bibitem [{\citenamefont {MacKintosh}\ and\ \citenamefont
  {Levine}(2008)}]{MacKintosh:2008}%
  \BibitemOpen
  \bibfield  {author} {\bibinfo {author} {\bibfnamefont {F.~C.}\ \bibnamefont
  {MacKintosh}}\ and\ \bibinfo {author} {\bibfnamefont {A.~J.}\ \bibnamefont
  {Levine}},\ }\bibfield  {title} {\bibinfo {title} {Nonequilibrium mechanics
  and dynamics of motor-activated gels},\ }\href
  {https://doi.org/10.1103/PhysRevLett.100.018104} {\bibfield  {journal}
  {\bibinfo  {journal} {Phys. Rev. Lett.}\ }\textbf {\bibinfo {volume} {100}},\
  \bibinfo {pages} {018104} (\bibinfo {year} {2008})}\BibitemShut {NoStop}%
\bibitem [{\citenamefont {Banerjee}\ and\ \citenamefont
  {Marchetti}(2011)}]{Banerjee:2011}%
  \BibitemOpen
  \bibfield  {author} {\bibinfo {author} {\bibfnamefont {S.}~\bibnamefont
  {Banerjee}}\ and\ \bibinfo {author} {\bibfnamefont {M.~C.}\ \bibnamefont
  {Marchetti}},\ }\bibfield  {title} {\bibinfo {title} {Instabilities and
  oscillations in isotropic active gels},\ }\href
  {https://doi.org/10.1039/C0SM00494D} {\bibfield  {journal} {\bibinfo
  {journal} {Soft Matter}\ }\textbf {\bibinfo {volume} {7}},\ \bibinfo {pages}
  {463} (\bibinfo {year} {2011})}\BibitemShut {NoStop}%
\bibitem [{\citenamefont {Ronceray}\ \emph {et~al.}(2016)\citenamefont
  {Ronceray}, \citenamefont {Broedersz},\ and\ \citenamefont
  {Lenz}}]{Ronceray:2016}%
  \BibitemOpen
  \bibfield  {author} {\bibinfo {author} {\bibfnamefont {P.}~\bibnamefont
  {Ronceray}}, \bibinfo {author} {\bibfnamefont {C.~P.}\ \bibnamefont
  {Broedersz}},\ and\ \bibinfo {author} {\bibfnamefont {M.}~\bibnamefont
  {Lenz}},\ }\bibfield  {title} {\bibinfo {title} {Fiber networks amplify
  active stress},\ }\href {https://doi.org/10.1073/pnas.1514208113} {\bibfield
  {journal} {\bibinfo  {journal} {Proceedings of the National Academy of
  Sciences}\ }\textbf {\bibinfo {volume} {113}},\ \bibinfo {pages} {2827}
  (\bibinfo {year} {2016})},\ \Eprint
  {https://arxiv.org/abs/https://www.pnas.org/doi/pdf/10.1073/pnas.1514208113}
  {https://www.pnas.org/doi/pdf/10.1073/pnas.1514208113} \BibitemShut {NoStop}%
\bibitem [{\citenamefont {Curatolo}\ \emph {et~al.}(2017)\citenamefont
  {Curatolo}, \citenamefont {Gabriele},\ and\ \citenamefont
  {Teresi}}]{Curatolo:2017_M}%
  \BibitemOpen
  \bibfield  {author} {\bibinfo {author} {\bibfnamefont {M.}~\bibnamefont
  {Curatolo}}, \bibinfo {author} {\bibfnamefont {S.}~\bibnamefont {Gabriele}},\
  and\ \bibinfo {author} {\bibfnamefont {L.}~\bibnamefont {Teresi}},\
  }\bibfield  {title} {\bibinfo {title} {Swelling and growth: a constitutive
  framework for active solids},\ }\href
  {https://doi.org/10.1007/s11012-017-0629-x} {\bibfield  {journal} {\bibinfo
  {journal} {Meccanica}\ }\textbf {\bibinfo {volume} {52}},\ \bibinfo {pages}
  {3443} (\bibinfo {year} {2017})}\BibitemShut {NoStop}%
\bibitem [{\citenamefont {Bacca}\ \emph {et~al.}(2019)\citenamefont {Bacca},
  \citenamefont {Saleh},\ and\ \citenamefont {McMeeking}}]{Bacca:2019}%
  \BibitemOpen
  \bibfield  {author} {\bibinfo {author} {\bibfnamefont {M.}~\bibnamefont
  {Bacca}}, \bibinfo {author} {\bibfnamefont {O.~A.}\ \bibnamefont {Saleh}},\
  and\ \bibinfo {author} {\bibfnamefont {R.~M.}\ \bibnamefont {McMeeking}},\
  }\bibfield  {title} {\bibinfo {title} {Contraction of polymer gels created by
  the activity of molecular motors},\ }\href
  {https://doi.org/10.1039/C8SM02598C} {\bibfield  {journal} {\bibinfo
  {journal} {Soft Matter}\ }\textbf {\bibinfo {volume} {15}},\ \bibinfo {pages}
  {4467} (\bibinfo {year} {2019})}\BibitemShut {NoStop}%
\bibitem [{\citenamefont {Curatolo}\ \emph {et~al.}(2020)\citenamefont
  {Curatolo}, \citenamefont {Nardinocchi},\ and\ \citenamefont
  {Teresi}}]{Curatolo:2019}%
  \BibitemOpen
  \bibfield  {author} {\bibinfo {author} {\bibfnamefont {M.}~\bibnamefont
  {Curatolo}}, \bibinfo {author} {\bibfnamefont {P.}~\bibnamefont
  {Nardinocchi}},\ and\ \bibinfo {author} {\bibfnamefont {L.}~\bibnamefont
  {Teresi}},\ }\bibfield  {title} {\bibinfo {title} {Dynamics of active
  swelling in contractile polymer gels},\ }\href
  {https://doi.org/https://doi.org/10.1016/j.jmps.2019.103807} {\bibfield
  {journal} {\bibinfo  {journal} {Journal of the Mechanics and Physics of
  Solids}\ }\textbf {\bibinfo {volume} {135}},\ \bibinfo {pages} {103807}
  (\bibinfo {year} {2020})}\BibitemShut {NoStop}%
\bibitem [{\citenamefont {Curatolo}\ \emph {et~al.}(2021)\citenamefont
  {Curatolo}, \citenamefont {Nardinocchi},\ and\ \citenamefont
  {Teresi}}]{Curatolo:2021}%
  \BibitemOpen
  \bibfield  {author} {\bibinfo {author} {\bibfnamefont {M.}~\bibnamefont
  {Curatolo}}, \bibinfo {author} {\bibfnamefont {P.}~\bibnamefont
  {Nardinocchi}},\ and\ \bibinfo {author} {\bibfnamefont {L.}~\bibnamefont
  {Teresi}},\ }\bibfield  {title} {\bibinfo {title} {Mechanics of active gel
  spheres under bulk contraction},\ }\href
  {https://doi.org/https://doi.org/10.1016/j.ijmecsci.2020.106147} {\bibfield
  {journal} {\bibinfo  {journal} {International Journal of Mechanical
  Sciences}\ }\textbf {\bibinfo {volume} {193}},\ \bibinfo {pages} {106147}
  (\bibinfo {year} {2021})}\BibitemShut {NoStop}%
\bibitem [{\citenamefont {Teresi}\ \emph {et~al.}(2022)\citenamefont {Teresi},
  \citenamefont {Curatolo},\ and\ \citenamefont {Nardinocchi}}]{Elsevier:2022}%
  \BibitemOpen
  \bibfield  {author} {\bibinfo {author} {\bibfnamefont {L.}~\bibnamefont
  {Teresi}}, \bibinfo {author} {\bibfnamefont {M.}~\bibnamefont {Curatolo}},\
  and\ \bibinfo {author} {\bibfnamefont {P.}~\bibnamefont {Nardinocchi}},\
  }\bibfield  {title} {\bibinfo {title} {Chapter 9 - active gel: A continuum
  physics perspective},\ }in\ \href
  {https://doi.org/https://doi.org/10.1016/B978-0-323-85740-6.00001-7} {\emph
  {\bibinfo {booktitle} {Modeling of Mass Transport Processes in Biological
  Media}}},\ \bibinfo {editor} {edited by\ \bibinfo {editor} {\bibfnamefont
  {S.}~\bibnamefont {Becker}}, \bibinfo {editor} {\bibfnamefont {A.~V.}\
  \bibnamefont {Kuznetsov}}, \bibinfo {editor} {\bibfnamefont {F.}~\bibnamefont
  {{de Monte}}}, \bibinfo {editor} {\bibfnamefont {G.}~\bibnamefont
  {Pontrelli}},\ and\ \bibinfo {editor} {\bibfnamefont {D.}~\bibnamefont
  {Zhao}}}\ (\bibinfo  {publisher} {Academic Press},\ \bibinfo {year} {2022})\
  pp.\ \bibinfo {pages} {287--309}\BibitemShut {NoStop}%
\bibitem [{Note1()}]{Note1}%
  \BibitemOpen
  \bibinfo {note} {See Ref.\protect \citenum {Turzi:2017}, where a similar
  point of view has been used to model active nematic gels.}\BibitemShut
  {Stop}%
\bibitem [{\citenamefont {Doi}(2009)}]{Doi:2009}%
  \BibitemOpen
  \bibfield  {author} {\bibinfo {author} {\bibfnamefont {M.}~\bibnamefont
  {Doi}},\ }\bibfield  {title} {\bibinfo {title} {Gel dynamics},\ }\href
  {https://doi.org/10.1143/JPSJ.78.052001} {\bibfield  {journal} {\bibinfo
  {journal} {J. Phys. Soc. Jpn.}\ }\textbf {\bibinfo {volume} {78}},\ \bibinfo
  {pages} {052001} (\bibinfo {year} {2009})}\BibitemShut {NoStop}%
\bibitem [{\citenamefont {Chester}\ and\ \citenamefont
  {Anand}(2010)}]{Chester:2010}%
  \BibitemOpen
  \bibfield  {author} {\bibinfo {author} {\bibfnamefont {S.~A.}\ \bibnamefont
  {Chester}}\ and\ \bibinfo {author} {\bibfnamefont {L.}~\bibnamefont
  {Anand}},\ }\bibfield  {title} {\bibinfo {title} {A coupled theory of fluid
  permeation and large deformations for elastomeric materials},\ }\href
  {https://doi.org/http://dx.doi.org/10.1016/j.jmps.2010.07.020} {\bibfield
  {journal} {\bibinfo  {journal} {Journal of the Mechanics and Physics of
  Solids}\ }\textbf {\bibinfo {volume} {58}},\ \bibinfo {pages} {1879 }
  (\bibinfo {year} {2010})}\BibitemShut {NoStop}%
\bibitem [{\citenamefont {Lucantonio}\ \emph {et~al.}(2013)\citenamefont
  {Lucantonio}, \citenamefont {Nardinocchi},\ and\ \citenamefont
  {Teresi}}]{JMPS:2013}%
  \BibitemOpen
  \bibfield  {author} {\bibinfo {author} {\bibfnamefont {A.}~\bibnamefont
  {Lucantonio}}, \bibinfo {author} {\bibfnamefont {P.}~\bibnamefont
  {Nardinocchi}},\ and\ \bibinfo {author} {\bibfnamefont {L.}~\bibnamefont
  {Teresi}},\ }\bibfield  {title} {\bibinfo {title} {Transient analysis of
  swelling-induced large deformations in polymer gels},\ }\href
  {https://doi.org/10.1016/j.jmps.2012.07.010} {\bibfield  {journal} {\bibinfo
  {journal} {Journal of the Mechanics and Physics of Solids}\ }\textbf
  {\bibinfo {volume} {61}},\ \bibinfo {pages} {205 } (\bibinfo {year}
  {2013})}\BibitemShut {NoStop}%
\bibitem [{\citenamefont {Fujine}\ \emph {et~al.}(2015)\citenamefont {Fujine},
  \citenamefont {Takigawa},\ and\ \citenamefont {Urayama}}]{Fujine:2015}%
  \BibitemOpen
  \bibfield  {author} {\bibinfo {author} {\bibfnamefont {M.}~\bibnamefont
  {Fujine}}, \bibinfo {author} {\bibfnamefont {T.}~\bibnamefont {Takigawa}},\
  and\ \bibinfo {author} {\bibfnamefont {K.}~\bibnamefont {Urayama}},\
  }\bibfield  {title} {\bibinfo {title} {Strain-driven swelling and
  accompanying stress reduction in polymer gels under biaxial stretching},\
  }\href {https://doi.org/10.1021/acs.macromol.5b00642} {\bibfield  {journal}
  {\bibinfo  {journal} {Macromolecules}\ }\textbf {\bibinfo {volume} {48}},\
  \bibinfo {pages} {3622} (\bibinfo {year} {2015})},\ \Eprint
  {https://arxiv.org/abs/https://doi.org/10.1021/acs.macromol.5b00642}
  {https://doi.org/10.1021/acs.macromol.5b00642} \BibitemShut {NoStop}%
\bibitem [{\citenamefont {Curatolo}\ \emph {et~al.}(2018)\citenamefont
  {Curatolo}, \citenamefont {Nardinocchi},\ and\ \citenamefont
  {Teresi}}]{Curatolo:2018_SM}%
  \BibitemOpen
  \bibfield  {author} {\bibinfo {author} {\bibfnamefont {M.}~\bibnamefont
  {Curatolo}}, \bibinfo {author} {\bibfnamefont {P.}~\bibnamefont
  {Nardinocchi}},\ and\ \bibinfo {author} {\bibfnamefont {L.}~\bibnamefont
  {Teresi}},\ }\bibfield  {title} {\bibinfo {title} {Driving water cavitation
  in a hydrogel cavity},\ }\href {https://doi.org/10.1039/C8SM00100F}
  {\bibfield  {journal} {\bibinfo  {journal} {Soft Matter}\ }\textbf {\bibinfo
  {volume} {14}},\ \bibinfo {pages} {2310} (\bibinfo {year}
  {2018})}\BibitemShut {NoStop}%
\bibitem [{\citenamefont {Prost}\ \emph {et~al.}(2015)\citenamefont {Prost},
  \citenamefont {Julicher},\ and\ \citenamefont {Joanny}}]{Prost:2015}%
  \BibitemOpen
  \bibfield  {author} {\bibinfo {author} {\bibfnamefont {J.}~\bibnamefont
  {Prost}}, \bibinfo {author} {\bibfnamefont {F.}~\bibnamefont {Julicher}},\
  and\ \bibinfo {author} {\bibfnamefont {J.-F.}\ \bibnamefont {Joanny}},\
  }\bibfield  {title} {\bibinfo {title} {Active gel physics},\ }\href
  {https://doi.org/http://dx.doi.org/10.1016/j.ijnonlinmec.2014.05.007}
  {\bibfield  {journal} {\bibinfo  {journal} {Nature Physics}\ }\textbf
  {\bibinfo {volume} {11}},\ \bibinfo {pages} {111 } (\bibinfo {year}
  {2015})},\ \bibinfo {note} {mechanics of Rubber - in Memory of Alan
  Gent}\BibitemShut {NoStop}%
\bibitem [{\citenamefont {Gurtin}\ \emph {et~al.}(2010)\citenamefont {Gurtin},
  \citenamefont {Fried},\ and\ \citenamefont {Anand}}]{Gurtin:2010}%
  \BibitemOpen
  \bibfield  {author} {\bibinfo {author} {\bibfnamefont {M.}~\bibnamefont
  {Gurtin}}, \bibinfo {author} {\bibfnamefont {E.}~\bibnamefont {Fried}},\ and\
  \bibinfo {author} {\bibfnamefont {L.}~\bibnamefont {Anand}},\ }\href@noop {}
  {\emph {\bibinfo {title} {The Mechanics and Thermodynamics of Continua}}}\
  (\bibinfo  {publisher} {Cambridge University Press},\ \bibinfo {year}
  {2010})\BibitemShut {NoStop}%
\bibitem [{\citenamefont {Hong}\ \emph {et~al.}(2008)\citenamefont {Hong},
  \citenamefont {Zhao}, \citenamefont {Zhou},\ and\ \citenamefont
  {Suo}}]{Hong:2008}%
  \BibitemOpen
  \bibfield  {author} {\bibinfo {author} {\bibfnamefont {W.}~\bibnamefont
  {Hong}}, \bibinfo {author} {\bibfnamefont {X.}~\bibnamefont {Zhao}}, \bibinfo
  {author} {\bibfnamefont {J.}~\bibnamefont {Zhou}},\ and\ \bibinfo {author}
  {\bibfnamefont {Z.}~\bibnamefont {Suo}},\ }\bibfield  {title} {\bibinfo
  {title} {A theory of coupled diffusion and large deformation in polymeric
  gels},\ }\href {https://doi.org/http://doi.org/10.1016/j.jmps.2007.11.010}
  {\bibfield  {journal} {\bibinfo  {journal} {Journal of the Mechanics and
  Physics of Solids}\ }\textbf {\bibinfo {volume} {56}},\ \bibinfo {pages}
  {1779 } (\bibinfo {year} {2008})}\BibitemShut {NoStop}%
\bibitem [{Note2()}]{Note2}%
  \BibitemOpen
  \bibinfo {note} {See \cite {Elsevier:2022} for a detailed derivation of the
  equations below.}\BibitemShut {Stop}%
\bibitem [{\citenamefont {Flory}\ and\ \citenamefont
  {Rehner}(1943{\natexlab{a}})}]{FR1}%
  \BibitemOpen
  \bibfield  {author} {\bibinfo {author} {\bibfnamefont {P.~J.}\ \bibnamefont
  {Flory}}\ and\ \bibinfo {author} {\bibfnamefont {J.}~\bibnamefont {Rehner}},\
  }\bibfield  {title} {\bibinfo {title} {Statistical mechanics of cross-linked
  polymer networks i. rubberlike elasticity},\ }\href
  {https://doi.org/10.1063/1.1723791} {\bibfield  {journal} {\bibinfo
  {journal} {J Chem Phys}\ }\textbf {\bibinfo {volume} {11}},\ \bibinfo {pages}
  {512} (\bibinfo {year} {1943}{\natexlab{a}})}\BibitemShut {NoStop}%
\bibitem [{\citenamefont {Flory}\ and\ \citenamefont
  {Rehner}(1943{\natexlab{b}})}]{FR2}%
  \BibitemOpen
  \bibfield  {author} {\bibinfo {author} {\bibfnamefont {P.~J.}\ \bibnamefont
  {Flory}}\ and\ \bibinfo {author} {\bibfnamefont {J.}~\bibnamefont {Rehner}},\
  }\bibfield  {title} {\bibinfo {title} {Statistical mechanics of cross-linked
  polymer networks ii. swelling},\ }\href {https://doi.org/10.1063/1.1723792}
  {\bibfield  {journal} {\bibinfo  {journal} {J Chem Phys}\ }\textbf {\bibinfo
  {volume} {11}},\ \bibinfo {pages} {521} (\bibinfo {year}
  {1943}{\natexlab{b}})}\BibitemShut {NoStop}%
\bibitem [{\citenamefont {Haviv}\ \emph {et~al.}(2008)\citenamefont {Haviv},
  \citenamefont {Gillo}, \citenamefont {Backouche},\ and\ \citenamefont
  {Bernheim-Groswasser}}]{Haviv:2008}%
  \BibitemOpen
  \bibfield  {author} {\bibinfo {author} {\bibfnamefont {L.}~\bibnamefont
  {Haviv}}, \bibinfo {author} {\bibfnamefont {D.}~\bibnamefont {Gillo}},
  \bibinfo {author} {\bibfnamefont {F.}~\bibnamefont {Backouche}},\ and\
  \bibinfo {author} {\bibfnamefont {A.}~\bibnamefont {Bernheim-Groswasser}},\
  }\bibfield  {title} {\bibinfo {title} {A cytoskeletal demolition worker:
  Myosin ii acts as an actin depolymerization agent},\ }\href
  {https://doi.org/https://doi.org/10.1016/j.jmb.2007.09.066} {\bibfield
  {journal} {\bibinfo  {journal} {Journal of Molecular Biology}\ }\textbf
  {\bibinfo {volume} {375}},\ \bibinfo {pages} {325} (\bibinfo {year}
  {2008})}\BibitemShut {NoStop}%
\bibitem [{Note3()}]{Note3}%
  \BibitemOpen
  \bibinfo {note} {They have been got at Bernheim Lab at Ben Gurion University
  (Israel), following the same methods illustrated in \cite
  {Ideses:2018}.}\BibitemShut {Stop}%
\bibitem [{\citenamefont {Efrati}\ \emph {et~al.}(2009)\citenamefont {Efrati},
  \citenamefont {Sharon},\ and\ \citenamefont {Kupferman}}]{Efrati:2009}%
  \BibitemOpen
  \bibfield  {author} {\bibinfo {author} {\bibfnamefont {E.}~\bibnamefont
  {Efrati}}, \bibinfo {author} {\bibfnamefont {E.}~\bibnamefont {Sharon}},\
  and\ \bibinfo {author} {\bibfnamefont {R.}~\bibnamefont {Kupferman}},\
  }\bibfield  {title} {\bibinfo {title} {Buckling transition and boundary layer
  in non-euclidean plates},\ }\href
  {https://doi.org/10.1103/PhysRevE.80.016602} {\bibfield  {journal} {\bibinfo
  {journal} {Phys. Rev. E}\ }\textbf {\bibinfo {volume} {80}},\ \bibinfo
  {pages} {016602} (\bibinfo {year} {2009})}\BibitemShut {NoStop}%
\bibitem [{\citenamefont {Sharon}\ and\ \citenamefont
  {Efrati}(2010)}]{Sharon:2010}%
  \BibitemOpen
  \bibfield  {author} {\bibinfo {author} {\bibfnamefont {E.}~\bibnamefont
  {Sharon}}\ and\ \bibinfo {author} {\bibfnamefont {E.}~\bibnamefont
  {Efrati}},\ }\bibfield  {title} {\bibinfo {title} {The mechanics of
  non-euclidean plates},\ }\href {https://doi.org/10.1039/C0SM00479K}
  {\bibfield  {journal} {\bibinfo  {journal} {Soft Matter}\ }\textbf {\bibinfo
  {volume} {6}},\ \bibinfo {pages} {5693} (\bibinfo {year} {2010})}\BibitemShut
  {NoStop}%
\bibitem [{\citenamefont {Pezzulla}\ \emph {et~al.}(2015)\citenamefont
  {Pezzulla}, \citenamefont {Shillig}, \citenamefont {Nardinocchi},\ and\
  \citenamefont {Holmes}}]{Pezzulla:2015}%
  \BibitemOpen
  \bibfield  {author} {\bibinfo {author} {\bibfnamefont {M.}~\bibnamefont
  {Pezzulla}}, \bibinfo {author} {\bibfnamefont {S.~A.}\ \bibnamefont
  {Shillig}}, \bibinfo {author} {\bibfnamefont {P.}~\bibnamefont
  {Nardinocchi}},\ and\ \bibinfo {author} {\bibfnamefont {D.~P.}\ \bibnamefont
  {Holmes}},\ }\bibfield  {title} {\bibinfo {title} {Morphing of geometric
  composites via residual swelling},\ }\href
  {https://doi.org/10.1039/C5SM00863H} {\bibfield  {journal} {\bibinfo
  {journal} {Soft Matter}\ }\textbf {\bibinfo {volume} {11}},\ \bibinfo {pages}
  {5812} (\bibinfo {year} {2015})}\BibitemShut {NoStop}%
\bibitem [{\citenamefont {Turzi}(2017)}]{Turzi:2017}%
  \BibitemOpen
  \bibfield  {author} {\bibinfo {author} {\bibfnamefont {S.~S.}\ \bibnamefont
  {Turzi}},\ }\bibfield  {title} {\bibinfo {title} {Active nematic gels as
  active relaxing solids},\ }\href {https://doi.org/10.1103/PhysRevE.96.052603}
  {\bibfield  {journal} {\bibinfo  {journal} {Phys. Rev. E}\ }\textbf {\bibinfo
  {volume} {96}},\ \bibinfo {pages} {052603} (\bibinfo {year}
  {2017})}\BibitemShut {NoStop}%
\end{thebibliography}%

\end{document}